\documentclass[aps,prl,twocolumn,longbibliography,superscriptaddress]{revtex4-2}

\usepackage{comment}
\usepackage{amssymb}
\usepackage{amsbsy}
\usepackage{amsmath}
\usepackage{graphicx}
\usepackage{graphics}
\usepackage{setspace}
\usepackage{array}
\usepackage{color}
\usepackage{fontenc}
\usepackage{textcomp}

\usepackage{bm}
\usepackage{float}
\usepackage{pifont}

\usepackage[bookmarks=false,linkcolor=blue,urlcolor=blue,colorlinks,citecolor=blue]{hyperref}

\usepackage{soul}
\usepackage[T1]{fontenc}
\usepackage[percent]{overpic}

\begin{document}

\title{Exact Fractionalized Ground States in an Extended Spin-1 Kitaev Chain}
\author{Alwyn Jose Raja}
\email{alwynjoseraja2000@gmail.com}
\affiliation{Department of Physics, Indian Institute of Technology Madras, Chennai 600036, India}

\author{R. Ganesh}
\email{r.ganesh@brocku.ca}
\affiliation{Department of Physics, Brock University, St. Catharines, Ontario L2S 3A1, Canada}	

\date{\today}

\begin{abstract}
    Inspired by the Affleck-Kennedy-Lieb-Tasaki (AKLT) model, we present exact solutions for a spin-1 chain with Kitaev-like couplings. We consider an expanded Kitaev model with bilinear and biquadratic terms. At an exactly solvable point, the Hamiltonian can be reexpressed as a sum of projection operators. Unlike the AKLT model where projectors act on total spin, we project the component of spin along the bond direction. This leads to exponential ground-state degeneracy, expressed in terms of fractionalized spin-$\frac{1}{2}$ objects. Each ground state can be expressed concisely as a matrix product state.  
    We construct a phase diagram by varying the relative strength of bilinear and biquadratic terms. The fractionalized states provide a qualitative picture for the spin-1 Kitaev model, yielding approximate forms for the ground state and low-lying excitations. 
\end{abstract}

\maketitle

\noindent{\color{blue} \it{Introduction}.}---
Exactly solvable models play a key role in the field of frustrated magnetism. Two prominent examples are the Affleck-Kennedy-Lieb-Tasaki (AKLT) model\cite{AKLT1987} and the Kitaev model on the honeycomb lattice\cite{kitaev2006anyons}. In both models, an exact ground-state wavefunction is found by reexpressing the original spins in terms of fractionalized operators. These solutions have far-reaching utility. For example, the AKLT solution provides a qualitative picture for the ground state of the spin-1 Heisenberg chain\cite{Schollwock1996,Golinelli1999,affleck1988valence}. 
In this letter, we present an exactly solvable model that combines elements of the Kitaev and AKLT models. Invoking fractionalization, we describe an exponentially large family of ground states. In the process, we present ans\"atze for the ground state and lowest excitations of the spin-1 Kitaev chain \cite{sen2010spin,gordon2022insights,raja2025spin}.

\noindent{\color{blue} \it{Model}.}---We consider a chain of spin-$1$ moments with bilinear and biquadratic interactions of strengths $K$ and $Q$ respectively. Couplings alternate between X and Y character, as described by the Hamiltonian
\begin{eqnarray}
\nonumber \hat{H}_{\theta} ~=&~&\sum_j \left[ K \left\{
\hat{S}_{2j}^x \hat{S}_{2j+1}^x + 
\hat{S}_{2j+1}^y \hat{S}_{2j+2}^y 
\right\}\right. \\
&+& \left. Q\left\{
\left(\hat{S}_{2j}^x \hat{S}_{2j+1}^x\right)^2 + 
\left(\hat{S}_{2j+1}^y \hat{S}_{2j+2}^y \right)^2
\right\}\right],
\label{eq.H}
\end{eqnarray}
where $j$ runs over integers. 
Upto an overall energy scale, we reduce $K$ and $Q$ to a single variable $\theta$, with $K\equiv \cos \theta$ and $Q\equiv\sin\theta$.
We consider an $N$-spin chain with periodic boundary conditions, where $N$ is even. 
We present analytic forms for the ground state(s), supported by numerical exact diagonalization results for $N=4,6,8,10,12$. We have conserved quantities on every bond, 
\begin{eqnarray}
\nonumber \hat{W}_{2j} = e^{i\pi \hat{S}_{2j}^y}e^{i\pi \hat{S}_{2j+1}^y},\\
\hat{W}_{2j+1} = e^{i\pi \hat{S}_{2j+1}^x} e^{i \pi\hat{S}_{2j+2}^x},
\label{eq.Ws}
\end{eqnarray}
that commute with the Hamiltonian and with each other. As they square to unity, they take eigenvalues $\pm 1$. 

\noindent{\color{blue} \it{Exactly solvable point}.}---When $\theta=\frac{\pi}{4}$, the linear and bilinear couplings are positive and equal. The Hamiltonian can be written as a sum of projectors,
\begin{eqnarray}
 \hat{H}_{\theta=\frac{\pi}{4}} = \sqrt{2}\sum_{j} 
\left\{
\hat{P}_{\left(\hat{S}_{2j}^x + \hat{S}_{2j+1}^x =\pm 2\right)} + 
\hat{P}_{\left(\hat{S}_{2j+1}^y + \hat{S}_{2j+2}^y =\pm 2\right)} 
\right\}.~~~~
\label{eq.Hproject}
\end{eqnarray}
The $\hat{P}$ operators here project onto maximal values of angular momentum along the bond direction --- see Supplement\cite{supplementary}. This is in contrast to the AKLT Hamiltonian where we project onto maximal total spin. As in the AKLT model, the projection operators here do not commute on neighbouring bonds.

\begin{figure}
    \centering
    \includegraphics[width=\linewidth]{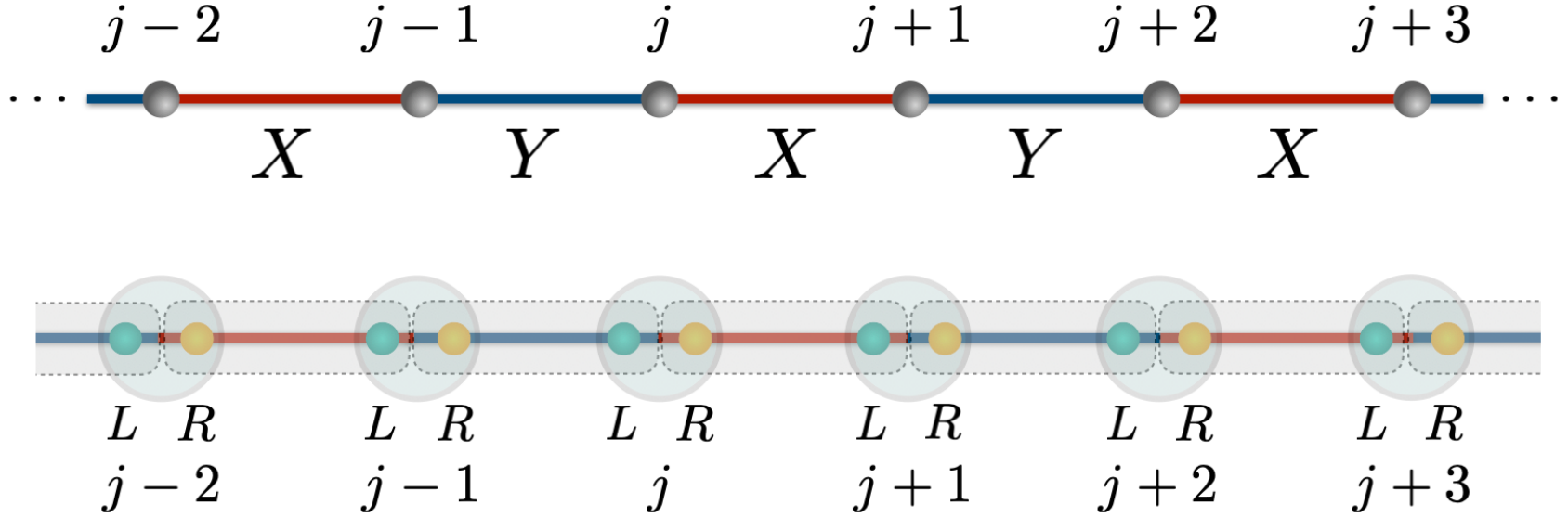}
    \caption{Fractionalized ground states at an exactly solvable point. Top: Bonds of the spin-1 Kitaev chain alternate between X and Y couplings. Bottom: The spin-1 moment at each site fractionalizes into two spin-$\frac{1}{2}$ objects or spinons, indicated as L (left) and R (right). On each bond, the spinons at the ends are placed in one of two valence bond wavefunctions. Finally, at every site, the L and R spinons are projected onto the local spin-1 space.}
    \label{fig.cartoon}
\end{figure}

Any state that is annihilated by all projection operators must necessarily be a ground state. We construct such states as follows. We fractionalize the spin-1 moment on each site into two spin-$\frac{1}{2}$ objects or spinons, `left' (L) and `right' (R). We may then write $\hat{S}_{j}^\lambda \rightarrow \hat{\sigma}_{j,L}^\lambda+\hat{\sigma}_{j,R}^\lambda$, where $\sigma$'s represent spin-$\frac{1}{2}$ operators and $\lambda=x,y$ or $z$. We next take pairs of spinons close to the centre of each bond and place them in a `bond-state'. The bond-states here are defined as:
\begin{itemize}
\item Singlet: $\vert s\rangle = \frac{1}{\sqrt{2}} \left\{\vert \! \uparrow \downarrow \rangle - \vert \!\downarrow \uparrow \rangle \right\}$,
\item Triplet-x: $\vert t_x \rangle = \frac{1}{\sqrt{2}} \left\{\vert \!\uparrow \uparrow \rangle - \vert \!\downarrow \downarrow \rangle \right\}$,
\item Triplet-y: $\vert t_y \rangle = \frac{i}{\sqrt{2}} \left\{\vert \!\uparrow \uparrow \rangle + \vert\! \downarrow \downarrow \rangle \right\}$.
\end{itemize}
The singlet state has net zero moment along any direction. The triplet-$\lambda$ state has net zero moment along direction $\lambda$, where $\lambda=x,y$. We now form bond-states according to Fig.~\ref{fig.cartoon}. On each X bond, we have a two-fold choice between a singlet and a triplet-x, both of which have zero net moment in the spin-x direction. 
The net bond-spin along X is given by
\begin{eqnarray}
    \hat{S}_{j}^x+\hat{S}_{j+1}^x = \hat{\sigma}_{j,L}^x+\{\hat{\sigma}_{j,R}^x+\hat{\sigma}_{j+1,L}^x\} +  \hat{\sigma}_{j+1,R}^x.
\end{eqnarray}
The term in braces vanishes as $(j,R)$ and $(j+1,L)$ form a bond-state with vanishing moment along the x-direction. We are left with contributions from two spin-$\frac{1}{2}$ moments that are away from the bond centre. As a result, the net spin along X can be -1, 0, or +1, but not $\pm 2$. This state is annihilated by the projection operator
$\hat{P}_{\left(\hat{S}_{j}^x +\hat{S}_{j+1}^x=\pm2\right)}$ in the Hamiltonian. 
Similarly, on every Y bond, we place either a singlet or a triplet-y. This satisfies the projection operators along y in the Hamiltonian.

With $N$ bonds and periodic boundaries, we arrive at $2^{N}$ fractionalized wavefunctions. To obtain physical states, we project onto the spin-1 sector at every site. Of the resulting $2^{N}$ wavefunctions, one is the AKLT state obtained by choosing a singlet bond-state for each bond. 
After projection, the fractionalized states are no longer orthogonal to one another. We later orthogonalize them to form a set of $2^N$ distinct states.

\begin{figure}
    \centering
    \includegraphics[width=\linewidth]{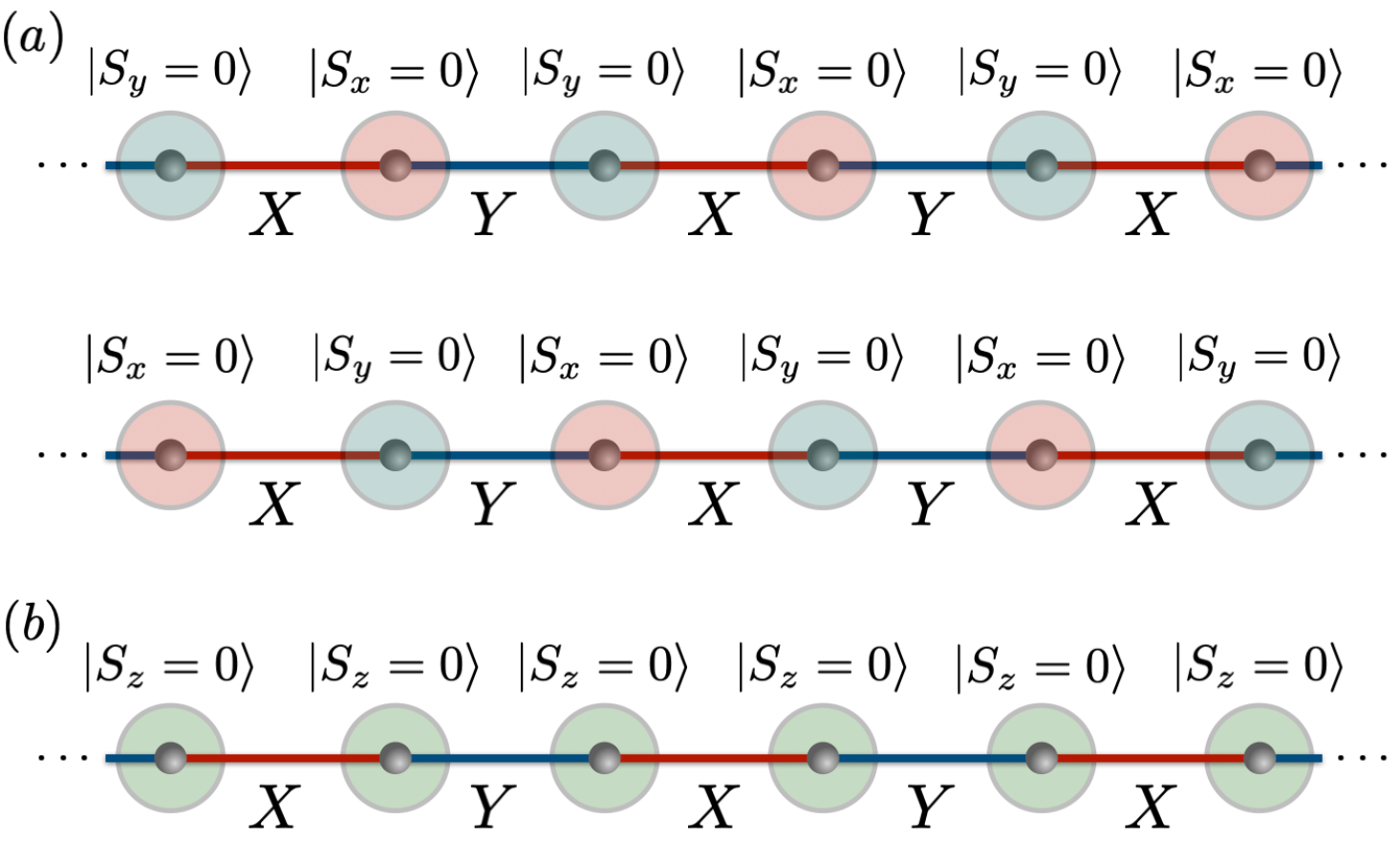}
    \caption{
    Direct-product ground states. (a) Spins are placed in $\vert S_x = 0\rangle$ and $\vert S_y = 0\rangle$ states in an alternating fashion. These two states are ground states when $Q$ is positive and $Q\geq \vert K\vert$.
    (b) All spins are placed in $\vert S_z = 0\rangle$ state, yielding the ground state when $K=0$ and $Q<0$.}
    \label{fig.directproduct}
\end{figure}

\noindent{\color{blue} \it{Additional ground state}.}---
At the exactly solvable $\theta=\frac{\pi}{4}$ point where the Hamiltonian is a sum of projectors, the ground-state energy is zero. Apart from the fractionalized construction, we may write two direct-product states with the same energy. As depicted in Fig.~\ref{fig.directproduct}(a), we place the spin-1 moment at each site in $|S_x=0\rangle$ or $|S_y=0\rangle$ states in an alternating fashion. The two resulting wavefunctions are readily seen to be eigenstates of the Hamiltonian in Eq.~\ref{eq.H}. Regardless of the values of $\theta$, they have eigenvalue zero. When $\theta=\frac{\pi}{4}$, they are degenerate with the fractionalized ground states described above. 

We make two assertions regarding the direct-product states of Fig.~\ref{fig.directproduct}(a): (i) their symmetric linear combination is, in fact, a linear superposition of fractionalized states. (ii) Their antisymmetric combination is orthogonal to all fractionalized states. It provides a distinct ground state at $\theta=\frac{\pi}{4}$, taking the ground state degeneracy to $2^{N}+1$. We substantiate these assertions in the Supplement\cite{supplementary}.

\begin{figure}
    \centering
    \includegraphics[width=\linewidth]{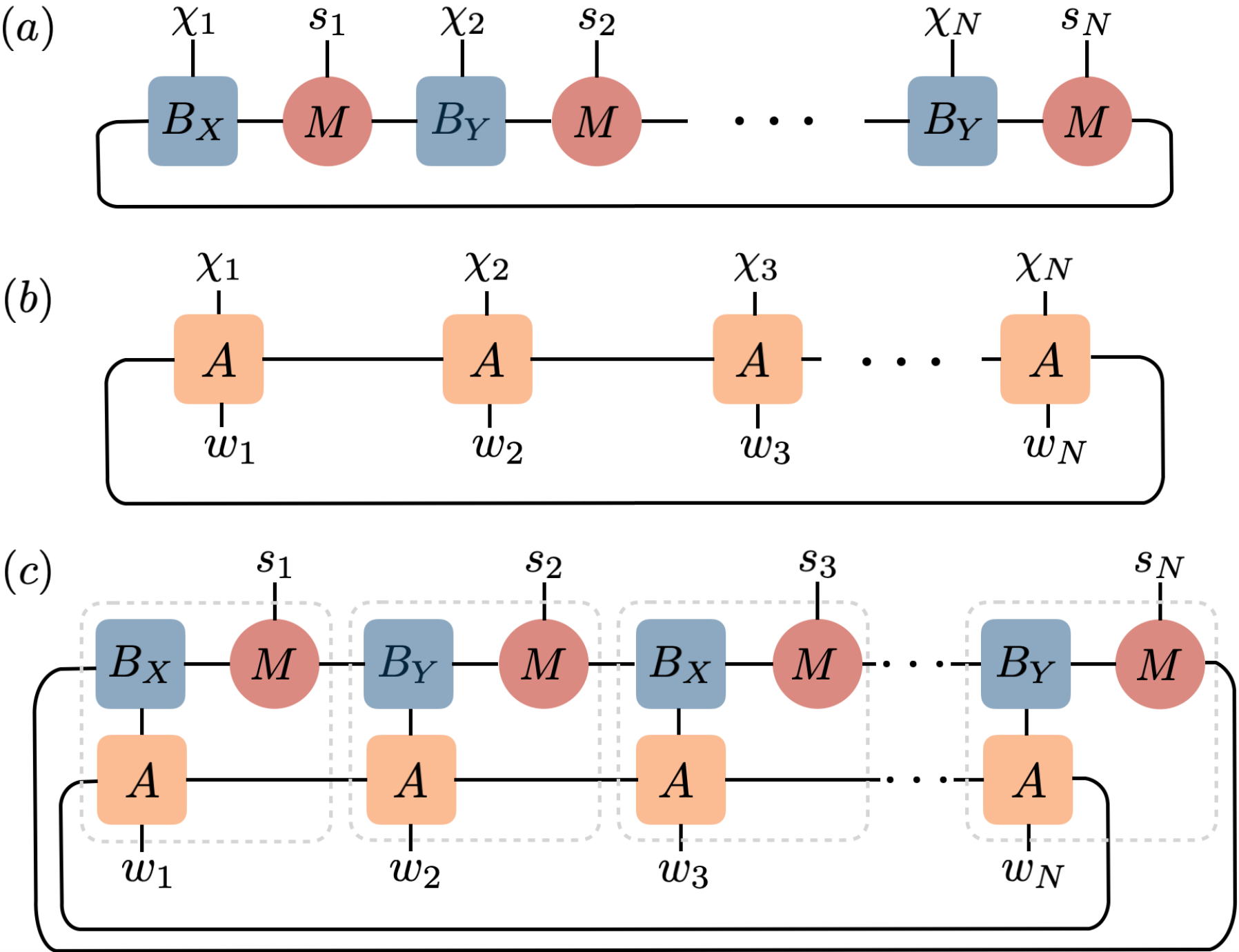}
    \caption{Matrix product state representations. (a) Fractionalized ground states at the exactly solvable point with $\theta=\frac{\pi}{4}$. The wavefunctions are not orthogonal to one another. Indices $\{ \chi\}$ represent the bond-state at each bond, while $\{ s\}$ represent the physical spin quantum numbers, $s_j=-1,0,1$. 
    (b) Ground state of the generalized cluster model. Indices $\{ w\}$ represent bond-conserved quantities. 
    (c) Orthogonalized ground states at the exactly solvable point, obtained by contracting MPSs from (a) and (b). Matrices enclosed within dashed lines can each be contracted into a single matrix of bond dimension four. Wavefunctions shown here are unnormalized. }
    \label{fig.MPSs}
\end{figure}

\noindent{\color{blue} \it{Matrix product state representation}.}---It is well known that the AKLT state can be concisely expressed as a matrix product state (MPS)\cite{schollwock2011density,Cirac2021,Tasaki2020}. This description can be adapted to describe all fractionalized ground states of Fig.~\ref{fig.cartoon}.
We introduce variables on each bond: $\chi_j=\uparrow,\downarrow$ on the bond connecting site $j$ to site $j+1$. The singlet bond-state is represented as $\chi=\uparrow$, while the appropriate triplet ($\vert t_x \rangle$ for an X bond and $\vert t_y\rangle$ for a Y bond) is denoted as $\chi=\downarrow$. 
A fractionalized ground state for a system with $N$ bonds and periodic boundaries is represented as $\vert \chi_1,\ldots,\chi_N \rangle$.
To construct an MPS, we define
\begin{eqnarray}
    B_X^\uparrow &=& B_Y^\uparrow =\left(
    \begin{array}{cc}
     0 & \frac{1}{\sqrt{2}} \\
     -\frac{1}{\sqrt{2}} & 0
    \end{array}
    \right),\nonumber\\
    B_X^\downarrow &=& \left(
    \begin{array}{cc}
     \frac{1}{\sqrt{2}} & 0\\
     0 & -\frac{1}{\sqrt{2}}
    \end{array}
    \right);~~    B_Y^\downarrow = \left(
    \begin{array}{cc}
    \frac{i}{\sqrt{2}} & 0 \\
     0 & \frac{i}{\sqrt{2}}
    \end{array}
    \right).~~~
    \label{eq.XYbondstates}
\end{eqnarray}
Matrices $B_X^\uparrow$ and $B_Y^\uparrow$ encode bond-state $\vert s \rangle$, while $B_X^\downarrow$ and $B_Y^\downarrow$ encode $\vert t_x \rangle$ and $\vert t_y\rangle$ respectively. We next define matrices that project onto the $s=-1,0,+1$ levels of a spin-1 moment,
\begin{eqnarray}
    M^{+1} = \left(
    \begin{array}{cc}
     1 & 0 \\
     0 & 0
    \end{array}
    \right);
    M^{-1} = \left(
    \begin{array}{cc}
     0 & 0\\
     0 & 1
    \end{array}
    \right);    M^{0} = \left(
    \begin{array}{cc}
    0 & \frac{1}{\sqrt{2}} \\
     \frac{1}{\sqrt{2}} & 0
    \end{array}
    \right)\! .~~~~
\end{eqnarray}
Using these matrices, a fractionalized state $\vert \chi_1,\ldots,\chi_N\rangle$ takes the form of the MPS shown in Fig.~\ref{fig.MPSs}(a). Crucially, MPS states corresponding to distinct $\{ \chi \}$'s are not orthogonal to one another.

\noindent{\color{blue} \it{Role of conserved quantities}.}---We use the conserved quantities of Eq.~\ref{eq.Ws} to orthogonalize the fractionalized ground states. We first examine the action of $\hat{W}$ operators on a fractionalized state denoted as $\vert  \chi_1,\ldots,\chi_N\rangle$. As shown in the supplement\cite{supplementary}, we find
\begin{eqnarray}
\nonumber    &\hat{W}_{k}\vert  \chi_1,\ldots,\chi_k,\ldots,\chi_N\rangle \\&= g(\chi_k)
    \vert  \ldots, {\chi}_{k-2},\bar{\chi}_{k-1},\chi_k,\bar{\chi}_{k+1},{\chi}_{k+2},\ldots\rangle,~~
    \label{eq.Waction}
\end{eqnarray}
where $g(\uparrow)=+1$  and $g(\downarrow)=-1$. The overbar ($\bar{\chi}$) represents a spin flip. The $\hat{W}_{k}$ operator flips the bond variables on bonds adjacent to the reference bond, leaving all others unchanged. Based on this relation, we express $\hat{W}_{k}$ (acting within the space of fractionalized states) as
\begin{eqnarray}
  \hat{W}_{k} = \sigma^x_{k-1}\sigma^z_{k}\sigma^x_{k+1},  
\end{eqnarray}
where $\sigma$'s are Pauli matrices that act on the $\chi=\uparrow,\downarrow$ degrees of freedom. We next seek superpositions of fractionalized states with specific values of the bond conserved quantities. We define a `selection Hamiltonian',
\begin{eqnarray}
    \hat{H}_{\{w_1,\ldots,w_N\}}= - \sum_{k=1}^N w_k ~\sigma^x_{k-1}\sigma^z_{k}\sigma^x_{k+1}.
    \label{eq.Hwpert}
\end{eqnarray}
Here, $\{w_1,\ldots,w_N\}$ represent desired eigenvalues of the bond conserved quantities defined in Eq.~\ref{eq.Ws}. 
Starting from the degenerate limit where all fractionalized states have the same energy, 
the selection Hamiltonian lowers the energy of a state with the desired $\{ w \}$ values.   
Remarkably, Eq.~\ref{eq.Hwpert} is a generalized form of the well-known cluster model \cite{Tasaki2020,Wei2022,suzuki1971relationship,raussendorf2001one, keating2004random,son2011quantum, PhysRevB.96.165124, yanagihara2020exact, sadaf2025quantum} that can be solved exactly, see Supplement\cite{supplementary}. The ground state of Eq.~\ref{eq.Hwpert} can be written as the MPS depicted in Fig.~\ref{fig.MPSs}(b), where 
\begin{eqnarray}
    A^{\chi=\uparrow}_{w=+1}=A^{\chi=\downarrow}_{w=-1}=\frac{1}{\sqrt{2}}\left(
    \begin{array}{cc}
    1 & 1 \\
    1 & -1 \end{array}\right), \nonumber \\
    A^{\chi=\downarrow}_{w=+1}=A^{\chi=\uparrow}_{w=-1}=\frac{1}{\sqrt{2}}
    \left(
    \begin{array}{cc}
        1 & 1 \\
        -1 & 1
    \end{array}\right).
    \label{eq.A_matrices}
\end{eqnarray}

Here, each $w$ index can take values $+1$ or $-1$. $\chi$'s represent `physical' indices. They are, in fact, bond variables that define a fractionalized state as given by Fig.~\ref{fig.MPSs}(a). To express the desired $\{w_1,\ldots,w_N\}$ state in terms of the original spin-1 moments, we contract the MPSs of Figs.~\ref{fig.MPSs}(a) and (b) to obtain Fig.~\ref{fig.MPSs}(c). The $\{s\}$ indices represent spin-1 states at each site, while $\{ w\}$ indices represent bond-conserved quantities. 

We draw three conclusions from the MPS representation of Fig.~\ref{fig.MPSs}(c): (i) We have $2^N$ legitimate wavefunctions, one for each choice of $\{w_1,\ldots,w_N\}$. Each is a linear superposition of the $2^N$ fractionalized states of Fig.~\ref{fig.cartoon}. 
(ii) They are MPSs of bond dimension four. In Fig.~\ref{fig.MPSs}(c),  each set of three matrices (shown within a box) can be combined into a single $4\times 4$ matrix, see Supplement\cite{supplementary}. (iii) Wavefunctions for distinct choices of $\{ w_1,\ldots,w_N\}$ are orthogonal, as they differ in conserved quantities.

\noindent{\color{blue} \it{Phase diagram}.}---
Having discussed the exactly solvable point with $\theta=\frac{\pi}{4}$, we next map the ground state phase diagram as a function of $\theta$. 
The Hamiltonian with parameters $(K,Q)$ maps to that at $(-K,Q)$ via a spin rotation (by $\pi$ about Z) at every other site. As a result, the phase diagram has a mirror reflection symmetry, with $\theta \equiv \pi-\theta$.  
Corresponding to $\theta=\frac{\pi}{4}$, an equivalent exactly solvable point exists at $\theta=\frac{3\pi}{4}$. 

\begin{figure}
    \centering
    \includegraphics[width=0.6\linewidth]{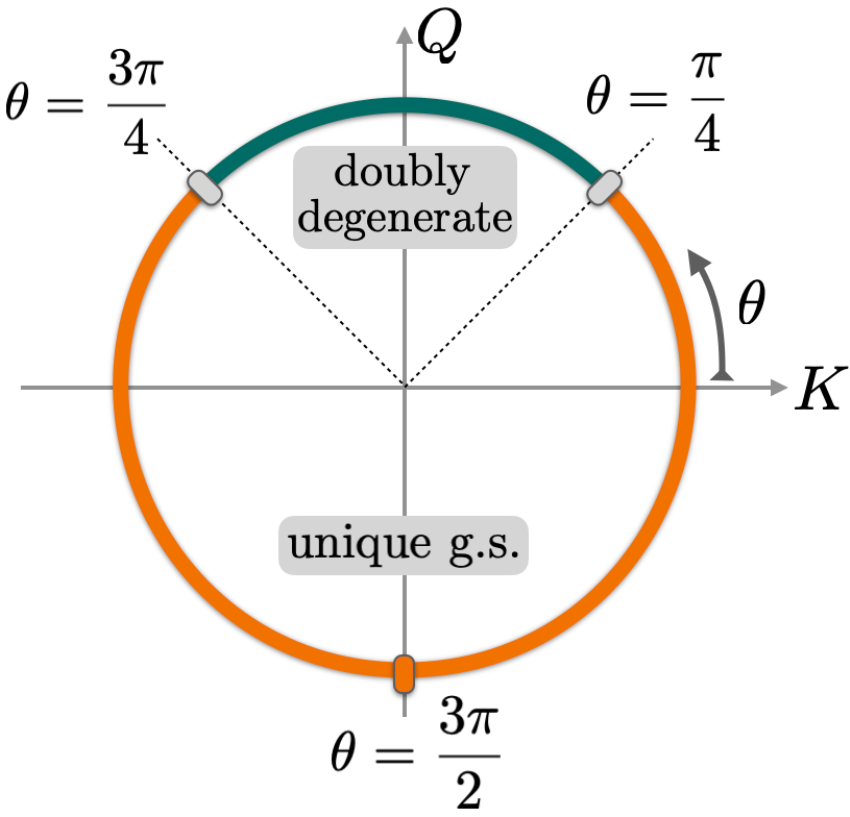}
    \caption{Ground state phase diagram of the spin-1 bilinear-biquadratic model. 
    In the `doubly degenerate' region, the ground states are the direct-product states of Fig.~\ref{fig.directproduct}(a).
    The two phases are separated by exactly solvable points at $\theta = \frac{\pi}{4}$ and $\frac{3\pi}{4}$. A special point appears at $\theta=\frac{3\pi}{2}$, with a unique direct-product ground state where all spins are in the $S_z=0$ state, as shown in Fig.~\ref{fig.directproduct}(b).
    }
    \label{fig.phasediagram}
\end{figure}

Based on exact diagonalization supported by analytic arguments, we find two phases as shown in Fig.~\ref{fig.phasediagram}. 

(i) For $\frac{\pi}{4} < \theta < \frac{3\pi}{4}$, ground state energy can be rigorously seen to be pinned at zero, see Supplement\cite{supplementary}.
We have a doubly degenerate ground state, consisting of the direct-product states in Fig.~\ref{fig.directproduct}(a). In both states, bond conserved quantities are uniformly $-1$. 

(ii) For $\theta < \frac{\pi}{4}$ and $\theta > \frac{3\pi}{4}$, we have a unique ground state with negative energy. The ground state takes a particularly simple form at $\theta=\frac{3\pi}{2}$, where we have purely biquadratic couplings. At this point, we have a direct-product ground state with all spins in the $S_z=0$ state, as shown in Fig.~\ref{fig.directproduct}(b). This state carries an energy of $-1$ per bond, saturating a lower bound on the ground state energy, see supplement\cite{supplementary}. Bond conserved quantities are uniformly $+1$. Moving away from this point with $K$ as a perturbation, bond conserved quantities remain pinned at $+1$\cite{supplementary}.  Crucially, the ground state of the spin-1 Kitaev model ($\theta=0$) is adiabatically connected to the direct-product ground state at $\theta=\frac{3\pi}{2}$.

The exactly solvable points at $\theta=\frac{\pi}{4}$ and $\frac{3\pi}{4}$ serve as phase boundaries, with an extensive ground state degeneracy of $2^{N}+1$ as previously described. 

\begin{figure}
\includegraphics[width=0.48\columnwidth]{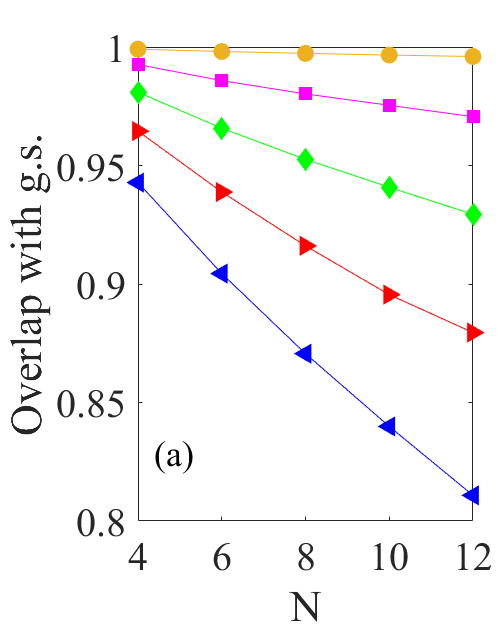}
\includegraphics[width=0.50\columnwidth]{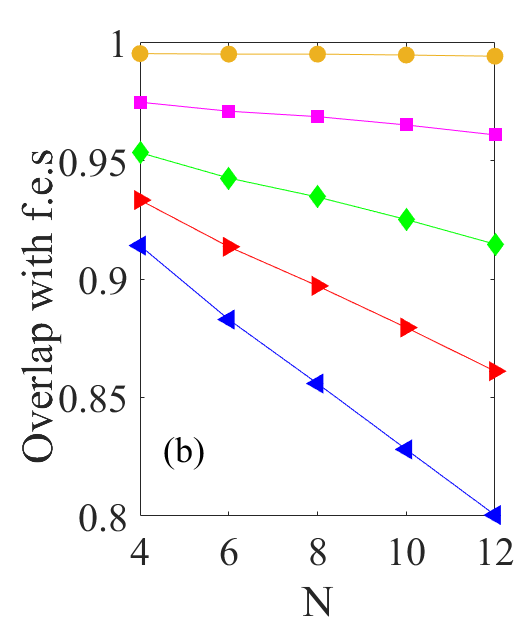}
\caption{(a) Overlap of the fractionalized state with all $w$'s set to $+1$ with the true ground state obtained from exact diagonalization. System size is plotted along the x-axis. The lines (from top to bottom) represent $\theta=40^\circ,30^\circ,20^\circ,10^\circ,0^\circ$. The lowest line in blue corresponds to the spin-1 Kitaev model with purely bilinear couplings. Overlaps approach unity as $\theta\rightarrow 45^\circ$. They decrease with increasing system size. (b) Overlap of a fractionalized state with the set of first excited states obtained from exact diagonalization. The fractionalized state is chosen with one $w$ set to -1 while all others are +1. }
\label{fig.overlaps}
\end{figure}

\noindent{\color{blue} \it{The spin-1 Kitaev model}.}---
At $\theta=\frac{\pi}{4}$, all fractionalized states of Fig.~\ref{fig.cartoon} are ground states. As $\theta$ decreases from $\frac{\pi}{4}$, one element of the fractionalized set is selected as the ground state. This state corresponds to choosing all $w$'s to be $+1$. It is obtained as an MPS from Fig.~\ref{fig.MPSs}(c) by setting all $w$'s to $+1$. By adiabatic continuity, we surmise that this state is a good approximation for the ground state when $\theta<\frac{\pi}{4}$. We examine this proposition in 
Fig.~\ref{fig.overlaps}(a), which plots the overlap of this ansatz with the true ground state obtained from exact diagonalization. As seen from the figure, the overlap is nearly unity when $\theta$ is just below $\frac{\pi}{4}$ and decreases as $\theta$ approaches zero. It diminishes with increasing system size, while retaining a reasonably strong magnitude. In particular, we obtain a substantial overlap for $\theta=0$. 
We conclude that the ground state of the spin-1 Kitaev chain is well-approximated by a fractionalized state with all $w$'s set to $+1$, resembling a ferromagnet in the $w$ variables.

We next consider the first excited state. We propose that the first excited state arises from a `spin-flip', where one of the $w$'s is flipped to $-1$. Indeed, exact diagonalization results show that the first excited state is $N$-fold degenerate -- matching the number of `single-spin-flip' states. In Fig.~\ref{fig.overlaps}(b), we plot the overlap of a single-spin-flip state (an MPS of Fig.~\ref{fig.MPSs}(c) with one $w$ set to -1, while all others are $+1$) with the space of first excited states obtained from exact diagonalization. We find reasonably strong overlap that decreases with system size and increases as we approach the exactly solvable point at $\theta=\frac{\pi}{4}$.

\begin{figure}

\includegraphics[width=\columnwidth]{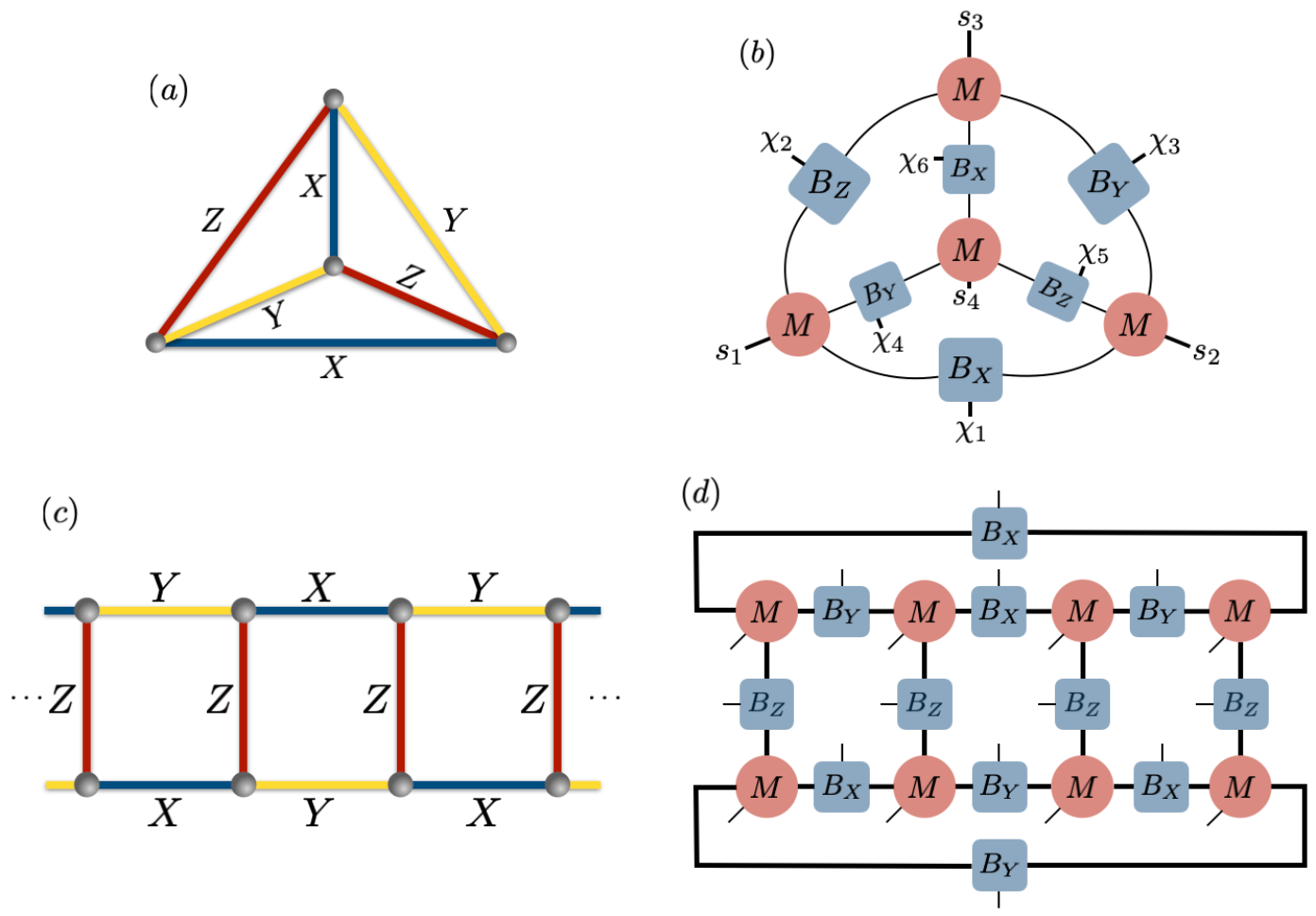}

\caption{(a) A spin-$\frac{3}{2}$ Kitaev-AKLT model on a tetrahedral cluster. The Hamiltonian is given by a sum of projectors onto $\hat{S}_{bond}^\alpha = \pm 3$, where $\alpha=x,y,z$ is the bond direction. (b) Fractionalized ground states at the exactly solvable Kitaev-AKLT point are constructed as PEPSs. The $\chi$ variables represent a two-fold choice on each bond, leading to 64 fractionalized wavefunctions. (c) A Kitaev-AKLT ladder with $S=\frac{3}{2}$ moments at each site. (d) PEPS representations for fractionalized ground states on a four-runged ladder with periodic boundaries. We have $2^{12}$ such states.
In (c) and (d), matrices $B_X$, $B_Y$ and $B_Z$ encode either a singlet or a bond-triplet state. The $M$ matrices represent projections onto the physical $S=\frac{3}{2}$ basis.
See Supplement\cite{supplementary} for explicit expressions.}
\label{fig.tetrahedron}
\end{figure}

\noindent{\color{blue} \it{Discussion}.}---
Our central result is that the spin-1 Kitaev chain lives in the shadow of an exactly solvable point with exponential ground-state degeneracy.
Exponential degeneracy with exact solvability is known in classical settings such as the triangular lattice Ising model and classical spin ice. 
Quantum problems with these features are rare. One example is the spin-$\frac{1}{2}$ Kitaev chain, integrable via a Jordan-Wigner transformation\cite{raja2025spin}. A second example occurs in three-colourable lattices with triangle motifs\cite{Changlani2018,Benton2021}.
In the Kitaev-AKLT model discussed here, these features arise from an AKLT-like construction with directional spin projectors. It is well known that the original AKLT idea generalizes to various lattices and spin quantum numbers\cite{affleck1988valence}, with the constraint that the coordination number must be a multiple of $2S$. The Kitaev-AKLT construction can also be extended in the same manner. Fig.~\ref{fig.tetrahedron} shows two examples, with spin-$\frac{3}{2}$ moments on a tetrahedral cluster as well as a ladder. At the exactly solvable Kitaev-AKLT point, 
we have $2^{N_b}$ fractionalized states as shown in Figs.~\ref{fig.tetrahedron}(b,d), where $N_b$ is the number of bonds. They can be viewed as projected entangled pair states (PEPS)\cite{Cirac2021}, rather than MPSs. We have checked that they form a linearly independent set of ground states. The precise form of the projector Hamiltonian is given in the Supplement\cite{supplementary}, along with ED results for comparison.

Exact MPS forms are of great current interest. AKLT and Majumdar-Ghosh models -- canonical examples of exactly solvable models -- admit MPS representations\cite{Cirac2021}. 
MPSs have been constructed for model quantum Hall wavefunctions\cite{Zaletel2012}. They have also been constructed in ferromagnetic models, with exponential degeneracy arising from rotational symmetry\cite{zhou2024exactmatrixproductstate,zhou2024entanglemententropytypescaleinvariant,zhou2024entanglemententropyonedimensionalflatband}. 
Ground states at a multicritical point in a spin-$\frac{1}{2}$ chain with anisotropic couplings have been shown to have MPS forms\cite{SaitoPRL2024}, with degeneracy that increases as the square of the system size. A prescription for constructing `frustration-free' Hamiltonians\cite{Tasaki2020}, such as Eq.~\ref{eq.Hproject} above, with exact MPS ground states has been put forward\cite{Saito2024}. Here, we provide exact MPS forms that encode bond-conserved quantities as quantum numbers that distinguish ground states.

The spin-1 Heisenberg bilinear-biquadratic model has been extensively studied\cite{Golinelli1999,Manmana2011}. Upon varying relative coupling strengths, this model connects the Takhtajan–Babujian\cite{Takhtajan1982,Babujian1982,Babujian1983} and the Uimin–Lai–Sutherland integrable points \cite{Uimin1970,Lai1974,Sutherland1975}. It contains the AKLT point and the Heisenberg chain as part of the Haldane phase, characterized by a string order parameter \cite{denNijs1989}. The Hamiltonian of Eq.~\ref{eq.H} represents a generalization of this model with Kitaev-like anisotropy. It is exactly solvable for the ground state(s) when $\theta=\frac{\pi}{4},\frac{3\pi}{4}$, $\frac{3\pi}{2}$ and for $\frac{\pi}{4}<\theta<\frac{3\pi}{4}$. 
For $\theta<\frac{\pi}{4}$ (and $>\frac{3\pi}{4}$), we obtain a regime analogous to the Haldane phase in the following respects.
It encompasses the Kitaev point at $\theta=0$, in analogy with the Haldane phase encompassing the Heisenberg chain.
With periodic boundaries, it exhibits a unique ground state. With open boundaries, the ground state is four-fold degenerate. This degeneracy can be precisely understood for $\theta=0$ (see Refs.~\cite{sen2010spin,raja2025spin}) and for $\theta=\frac{3\pi}{2}$ (see Supplement\cite{supplementary}). 
Unlike the Haldane phase, the $\theta<\frac{\pi}{4}$ phase is smoothly connected to a direct-product wavefunction -- the ground state at $\theta=\frac{3\pi}{2}$ -- and is therefore non-topological. This complements previous studies on entanglement properties of the spin-1 Kitaev model \cite{luo2021unusual}.
Future studies can examine whether a distinct string-like order parameter exists with Kitaev anisotropy\cite{you2020quantum,wei2025quantum}.

The Kitaev-AKLT model hosts an interesting phase transition at $\theta=\frac{\pi}{4}$, with similarities to the square-lattice Rokhsar-Kivelson (RK) model \cite{RKmodel1988,moessner2010quantum}. On either side of the transition, we have distinct ground states. The transition point has a projector Hamiltonian that gives rise to a large degeneracy. In the RK model, the degeneracy is related to the number of topological sectors. Here, it is exponential in system size. In both models, the transition is sharp; although quantum effects are fully taken into account, they do not smear the transition or produce an intermediate phase.

We have presented approximate MPS forms for low-lying states of the spin-1 Kitaev chain. An exciting future direction is to explore whether exact/approximate forms can be obtained for high-energy states as well \cite{caspers1982some,arovas1989two,PhysRevB.98.235155,PhysRevB.98.235156,PhysRevB.75.104411}. The fractionalized and direct-product states discussed here can serve as a variational basis set for related problems, e.g., with perturbations such as Heisenberg or $\Gamma$ couplings\cite{you2020quantum,luo2021unusual,wei2025quantum,pohle2023spin}.
 
\acknowledgements
    We thank Kirill Shtengel, Julien Vidal, Diptiman Sen and P\'eter Kr\'anitz for helpful discussions. RG thanks the National Sciences and Engineering Research Council of Canada (NSERC) for support through Discovery Grant 2022-05240, and the Office of Global Engagement, IIT Madras, for hospitality and support.
\vspace{-2mm}    
\bibliography{KitaevAKLT}
\newpage

\onecolumngrid 
\setcounter{secnumdepth}{3}
\renewcommand{\theequation}{S\arabic{equation}}
\renewcommand{\thefigure}{S\arabic{figure}}
\begin{center}
 \textbf{\large Supplemental Material for ``Exact Fractionalized Ground States in an Extended Spin-1 Kitaev Chain
''}\\[.5cm]
Alwyn Jose Raja$^1$ and R. Ganesh$^2$\\[.4cm]
{\itshape ${}^1$Department of Physics, Indian Institute of Technology Madras, Chennai 600036, India\\
\itshape ${}^2$Department of Physics, Brock University, St. Catharines, Ontario L2S 3A1, Canada\\}

(Dated: \today)\\[2cm]
\end{center}

\section{Convention for spin-1 matrices}
We use the standard definition of spin-1 operators,
\begin{eqnarray}
    \hat{S}_i^x=\frac{1}{\sqrt{2}}\begin{pmatrix}
        0 & 1 & 0\\
        1 & 0 & 1\\
        0 & 1 & 0
    \end{pmatrix}, ~~
    \hat{S}_i^y=\frac{1}{\sqrt{2}}\begin{pmatrix}
        0 & -i & 0\\
        i & 0 & -i\\
        0 & i & 0
    \end{pmatrix},~~
    \hat{S}_i^z=\begin{pmatrix}
        1 & 0 & 0\\
        0 & 0 & 0\\
        0 & 0 & -1
    \end{pmatrix},
\end{eqnarray}
with squared operators 
\begin{eqnarray}
    (\hat{S}_i^x)^2=\frac{1}{2}\begin{pmatrix}
        1 & 0 & 1\\
        0 & 2 & 0\\
        1 & 0 & 1
    \end{pmatrix}, ~~
    (\hat{S}_i^y)^2=\frac{1}{2}\begin{pmatrix}
        1 & 0 & -1\\
        0 & 2 & 0\\
        -1 & 0 & 1
    \end{pmatrix},~~
    (\hat{S}_i^z)^2=\begin{pmatrix}
        1 & 0 & 0\\
        0 & 0 & 0\\
        0 & 0 & 1
    \end{pmatrix}. 
\end{eqnarray}
These yield the well-known relations,
\begin{eqnarray}
    (\hat{S}^\alpha_i)^3=S^\alpha_i;~~
    \exp(i\theta \hat{S}_i^\alpha)=I+i\sin\theta \hat{S}_i^\alpha +(-1+\cos\theta)(\hat{S}_i^\alpha)^2;~~
    \exp(i\pi \hat{S}_i^\alpha)=I -2(\hat{S}_i^\alpha)^2,
\end{eqnarray}
where $\alpha=x,y,z$ and $I$ represents the $3\times 3 $ identity matrix. We have
\begin{eqnarray}
     \exp(i\pi \hat{S}_i^x)=-\begin{pmatrix}
        0 & 0 & 1\\
        0 & 1 & 0\\
        1 & 0 & 0
    \end{pmatrix},~~
    \exp(i\pi \hat{S}_i^y)=\begin{pmatrix}
        0 & 0 & 1\\
        0 & -1 & 0\\
        1 & 0 & 0
    \end{pmatrix},~~  
    \exp(i\pi \hat{S}_i^z)=\begin{pmatrix}
        -1 & 0 & 0\\
        0 & 1 & 0\\
        0 & 0 & -1
    \end{pmatrix}. 
\end{eqnarray}
With these definitions, it is easy to see that

\begin{eqnarray}
    \exp(i\pi \hat{S}_i^\alpha) \vert S_\beta=0 \rangle =(-1)^{\delta_{\alpha\beta}+1} \vert S_\beta=0 \rangle,
\end{eqnarray}
where $\alpha,\beta = x,y,z$. Here, $\vert S_\beta=0 \rangle$ are eigenvectors of $\hat{S}^\beta_i$ operator with eigenvalue 0.

\section{Projection onto maximal moments along the bond direction}

On a bond between sites $j$ and $j+1$, we consider the projection onto total-spin-x being $\pm 2$. We have
\begin{eqnarray}
   \hat{P}_{(\hat{S}_j^x + \hat{S}_{j+1}^x)=\pm2}
    =\frac{1}{12}
    \Big(
    (\hat{S}_j^x+\hat{S}_{j+1}^x)^4 - (\hat{S}_j^x+\hat{S}_{j+1}^x)^2
    \Big)
    =\frac{1}{2}\Big((\hat{S}_{j}^x \hat{S}_{j+1}^x) + (\hat{S}_{j}^x \hat{S}_{j+1}^x)^2\Big).
\end{eqnarray}
We have used properties of spin-1 operators described in the previous section, e.g., $\big(\hat{S}_j^x\big)^3=\hat{S}_j^x$. We may verify the final expression by explicitly checking all possible values: $\hat{S}_j^x=-1,0,+1$ and $\hat{S}_{j+1}^x=-1,0,+1$. On the same lines, we have 
$\hat{P}_{(\hat{S}_j^y + \hat{S}_{j+1}^y)=\pm2} = \frac{1}{2}\Big((S_{j}^y S_{j+1}^y) + (S_{j}^y S_{j+1}^y)^2\Big)$.

\section{Edge states {at $\theta=\frac{\pi}{4}$}}

\begin{figure}
    \centering
\includegraphics[width=0.5\linewidth]{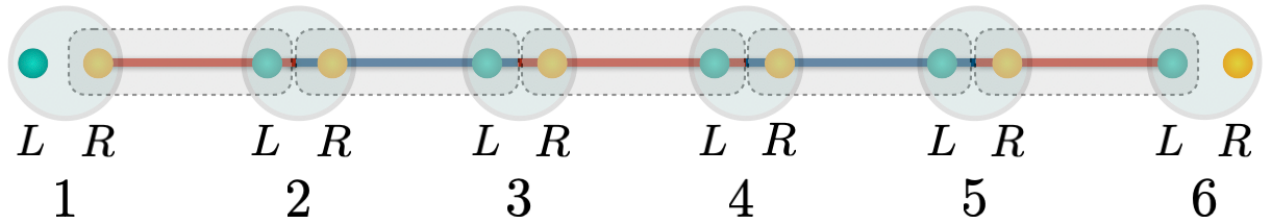}
    \caption{Open boundary conditions, illustrated for $N=6$. We have five bonds, each with a two-fold choice. We have dangling spinons at the edges. }
    \label{fig.edge}
\end{figure}

A key feature of the AKLT model is the appearance of an emergent, free spin-$\frac{1}{2}$ moment at each edge. We find a direct analogy at the exactly solvable $K=Q$ point here. In an open chain of $N$ (chosen to be even) spins with X-bonds at the two ends, exact diagonalization calculations show that the ground state degeneracy is $(2^{N+1}-1)$. This can be understood as follows. As illustrated in Fig.~\ref{fig.edge}, we have $N-1$ bonds. Each can be assigned one of two bond states as described in the main text. This leaves us with two dangling spinons at the edges. Naively, each spinon has two possible states, leading to four edge-configurations that are independent of the bulk. These arguments suggest a ground state degeneracy of $4\times 2^{N-1}$. However, one of these states turns out to be superfluous, i.e., linearly dependent on the others. 
This leaves us with a ground state degeneracy of $(2^{N+1}-1)$.

To illustrate this, we consider the simplest case of $N=2$. We have two sites with a single bond connecting them, say of the $Z$ type. We choose the bond to be of the $Z$ type so as to have easily recognizable wave functions. The arguments below can be modified for X or Y bonds by a global spin rotation. The Hamiltonian (corresponding to $\theta=\frac{\pi}{4}$) is given by 
\begin{equation}
     \hat{P}_{(\hat{S}^z_{1}+\hat{S}^z_{2})=\pm 2}=\frac{1}{2}\Big(\hat{S}^z_1 \hat{S}^z_2 + (\hat{S}^z_1)^2(\hat{S}^z_2)^2 \Big).
\end{equation}
The Hilbert space for this problem is 9-dimensional, arising from two spin-1 moments. Of the 9 states, 7 are ground states: i. $|1\rangle_1|0\rangle_2$, ii. $|0\rangle_1|1\rangle_2$, iii. $|1\rangle_1|-1\rangle_2$, iv. $|-1\rangle_1|1\rangle_2$, v. $|0\rangle_1|0\rangle_2$ vi. $|-1\rangle_1|0\rangle_2$, vii. $|0\rangle_1|-1\rangle_2$.
We have labelled states at each site as $|S=1,m_z\rangle \equiv |m_z\rangle$. States i and ii have $S^z_{total}=1$, iii-v have $S^z_{total}=0$, while the final two have $S^z_{total}=-1$.

We now switch to the fractionalized representation with two spinons per site, denoted as  $1_L,1_R,2_L,2_R$. To satisfy the projector Hamiltonian, we place $1_R$ and $2_L$ in a singlet $|s\rangle$ or triplet-z $|t_z\rangle$ bond state. The latter is defined as $\vert t_z\rangle = \frac{1}{\sqrt{2}}\{ \vert  \!  \uparrow \downarrow \rangle + \vert  \! \downarrow\uparrow\rangle \}$. Naively, the $1_L$ and $2_R$ spinons can each be in two states independently. This leads to eight possible states for the system. However, as explicitly shown above, the ground state degeneracy is seven. Thus, the naive construction contains one superfluous state. To see this, we project the eight fractionalized states onto the physical spin-1 space at each site:

\begin{enumerate}
    \item $\hat{P}_{1_L,1_R}^{S=1}~\hat{P}_{2_L,2_R}
    ^{S=1} \Big\{|\uparrow_{1L}\rangle|s_{1R,2L}\rangle|\uparrow_{2R}\rangle\Big\}=|1\rangle_1|0\rangle_2-|0\rangle_1|1\rangle_2$,
    \item $\hat{P}_{1_L,1_R}^{S=1}~\hat{P}_{2_L,2_R}^{S=1}\Big\{|\uparrow_{1L}\rangle|s_{1R,2L}\rangle|\downarrow_{2R}\rangle\Big\}=\frac{1}{\sqrt{2}}\Big(|1\rangle_1|-1\rangle_2-2|0\rangle_1|0\rangle_2\Big)$,
    \item $\hat{P}_{1_L,1_R}^{S=1}~\hat{P}_{2_L,2_R}^{S=1}\Big\{|\downarrow_{1L}\rangle|s_{1R,2L}\rangle|\uparrow_{2R}\rangle\Big\}=\frac{1}{\sqrt{2}}\Big(2|0\rangle_1|0\rangle_2-|-1\rangle_1|1\rangle_2\Big)$,
    \item $\hat{P}_{1_L,1_R}^{S=1}~\hat{P}_{2_L,2_R}^{S=1}\Big\{|\downarrow_{1L}\rangle|s_{1R,2L}\rangle|\downarrow_{2R}\rangle\Big\}=|0\rangle_1|-1\rangle_2-|-1\rangle_1|0\rangle_2$,
   \item$\hat{P}_{1_L,1_R}^{S=1}~\hat{P}_{2_L,2_R}^{S=1}\Big\{|\uparrow_{1L}\rangle|t_{z_{1R,2L}}\rangle|\uparrow_{2R}\rangle\Big\}=|1\rangle_1|0\rangle_2+|0\rangle_1|1\rangle_2$,
    \item $\hat{P}_{1_L,1_R}^{S=1}~\hat{P}_{2_L,2_R}^{S=1}\Big\{|\uparrow_{1L}\rangle|t_{z_{1R,2L}}\rangle|\downarrow_{2R}\rangle\Big\}=\frac{1}{\sqrt{2}}\Big(|1\rangle_1|-1\rangle_2+2|0\rangle_1|0\rangle_2\Big)$,
    \item $\hat{P}_{1_L,1_R}^{S=1}~\hat{P}_{2_L,2_R}^{S=1}\Big\{|\downarrow_{1L}\rangle|t_{z_{1R,2L}}\rangle|\uparrow_{2R}\rangle\Big\}=\frac{1}{\sqrt{2}}\Big(2|0\rangle_1|0\rangle_2+|-1\rangle_1|1\rangle_2\Big)$, 
    \item$\hat{P}_{1_L,1_R}^{S=1}~\hat{P}_{2_L,2_R}^{S=1}\Big\{|\downarrow_{1L}\rangle|t_{z_{1R,2L}}\rangle|\downarrow_{2R}\rangle\Big\}=|0\rangle_1|-1\rangle_2+|-1\rangle_1|0\rangle_2$. 
\end{enumerate}
Here, $\hat{P}^{S=1}$ is an operator that projects two spin-$\frac{1}{2}$ objects onto the physical spin-1 subspace. 
From these equations, we see that the eight states are not linearly independent, as
\begin{eqnarray}
\hat{P}_{1_L,1_R}^{S=1}~\hat{P}_{2_L,2_R}
    ^{S=1} \Big\{
    |\uparrow_{1L}\rangle|s_{1R,2L}\rangle|\downarrow_{2R}\rangle +
    |\downarrow_{1L}\rangle|s_{1R,2L}\rangle|\uparrow_{2R}\rangle +
    |\uparrow_{1L}\rangle|t_{z_{1R,2L}}\rangle|\downarrow_{2R}\rangle -
    |\downarrow_{1L}\rangle|t_{z_{1R,2L}}\rangle|\uparrow_{2R}\rangle
    \Big\}=0.~~
\end{eqnarray}
If one of the states is removed (e.g., $\vert \!\uparrow_{1L}\rangle|s_{1R,2L}\rangle|\downarrow_{2R}\rangle$), the remaining states have full rank. We find the same situation for any even $N$, with $(2^{N+1}-1)$ linearly independent states.

\section{Action of conserved operators}
We seek to demonstrate Eq.~7, starting from the definitions of $\hat{W}$ operators in Eq.~2 and the fractionalized construction. We first express a generic fractionalized state, following Fig.~1, as
\begin{eqnarray}
    \vert \chi_1,\chi_2,\ldots,\chi_N \rangle = \Big\{
        \hat{P}^{S=1}_1\otimes \hat{P}^{S=1}_2\otimes \ldots\otimes  \hat{P}^{S=1}_N 
    \Big\}~
    \Big\{
        \vert \chi_1 \rangle \otimes \vert \chi_2 \rangle \otimes \ldots \otimes \vert \chi_N \rangle\Big\}.
\end{eqnarray}
Here, $\vert \chi_j \rangle$ represents a bond state of spinons $\{j_R,(j+1)_L\}$
that is either a singlet or the appropriate triplet. The $\hat{P}^{S=1}$ operators are projectors onto the physical spin-1 space at each site. We next consider the action of $\hat{W}_j$ on this state,
\begin{eqnarray}
\nonumber     \hat{W}_j \vert \chi_1,\ldots,\chi_N\rangle &=& 
    \hat{W}_j \Big\{
        \hat{P}^{S=1}_1\otimes  \ldots\otimes  \hat{P}^{S=1}_N 
    \Big\}~
    \Big\{
        \vert \chi_1 \rangle \otimes \ldots \otimes \vert \chi_N \rangle\Big\}\\
      &=& \Big\{
        \hat{P}^{S=1}_1\otimes \ldots\otimes  \hat{P}^{S=1}_N 
    \Big\}~ \hat{W}_j ~
    \Big\{
        \vert \chi_1\rangle \otimes \ldots \otimes \vert \chi_N \rangle\Big\}.
        \label{eq.WPchi}
\end{eqnarray}  
We have used the fact that $\hat{W}$ operators commute with spin-1 projections at all sites. This is because a $\hat{W}$ operator implements a spin-rotation on two spins at the ends of a bond, without changing the total spin at either site. This can be seen explicitly as follows. We consider an X bond between sites $j$ and $j+1$, for concreteness. In the fractionalized construction, we introduce two spinons at each site. 

We express $\hat{W}_j$ in the expanded Hilbert space of the spin-$\frac{1}{2}$ partons following the convention of Fig.~1,
\begin{eqnarray}
    \hat{W}_{j}= \exp(i\pi\sigma_{j,L}^y)\otimes\exp(i\pi\sigma_{j,R}^y)\otimes\exp(i\pi\sigma_{j+1,L}^y)\otimes\exp(i\pi\sigma_{j+1,R}^y).
\label{eq.Wmat}
\end{eqnarray}
Here, $\sigma_y$'s are Pauli matrices that act on spin-$\frac{1}{2}$ moments.
In the expanded Hilbert space, the projection operator onto the spin-1 subspace at site $j$ is given by
\begin{equation}
    \hat{P}^{S=1}_{j}=\hat{P}^{S=1}_{j_L,j_R}=
    \begin{pmatrix}
        1 && 0 && 0 && 0\\ 
        0 && 1/2 && 1/2 && 0 \\
        0 && 1/2 && 1/2 && 0 \\
        0 && 0 && 0 && 1
    \end{pmatrix},
    \label{eq.Pmat}
\end{equation}
acting on $\begin{pmatrix}
    |\uparrow_{j,L}\rangle|\uparrow_{j,R}\rangle && |\uparrow_{j,L}\rangle|\downarrow_{j,R}\rangle & &|\downarrow_{j,L}\rangle|\uparrow_{j,R}\rangle &&|\downarrow_{j,L}\rangle|\downarrow_{j,R}\rangle
\end{pmatrix}^\dagger$. 
Using Eqs.~\ref{eq.Wmat} and \ref{eq.Pmat}, we can explicitly check that $[\hat{W}_{j},\hat{P}^{S=1}_{j_L,j_R}]=0$. In the same vein, we can check, by explicit construction, that $[\hat{W}_{j},\hat{P}^{S=1}_{(j+1)_L,(j+1)_R}]=0$. 
Projection operators at sites other than $j$ and $(j+1)$ trivially commute with $\hat{W}_j$. These arguments can also be checked to hold true on Y bonds. 
This justifies the last step in Eq.~\ref{eq.WPchi} where we have moved $\hat{W}_j$ past on-site projection operators.

Continuing from Eq.~\ref{eq.WPchi}, 

the $\hat{W}_{j}$ operator acts on bond states as follows,
\begin{eqnarray}
    \hat{W}_{j} ~\Big\{
    \ldots \otimes |s_{j-1}\rangle \otimes 
    |s_{j}\rangle \otimes 
    |s_{j+1}\rangle \otimes \ldots \Big\}
    &=&
    \Big\{
    \ldots \otimes|t_{j-1}\rangle \otimes 
    |s_{j}\rangle \otimes 
    |t_{j+1}\rangle
    \otimes \ldots \Big\}, \nonumber \\
    \hat{W}_{j} ~\Big\{
    \ldots \otimes |t_{j-1}\rangle \otimes 
    |s_{j}\rangle \otimes 
    |s_{j+1}\rangle \otimes \ldots \Big\}
    &=&
    \Big\{
    \ldots \otimes|s_{j-1}\rangle \otimes 
    |s_j\rangle \otimes 
    |t_{j+1}\rangle
    \otimes \ldots \Big\}, \nonumber\\
    \hat{W}_{j} ~\Big\{
    \ldots \otimes |s_{j-1}\rangle \otimes 
    |s_j\rangle \otimes 
    |t_{j+1}\rangle \otimes \ldots \Big\}
    &=&
    \Big\{
    \ldots \otimes|t_{j-1}\rangle \otimes 
    |s_j\rangle \otimes 
    |s_{j+1}\rangle
    \otimes \ldots \Big\}, \nonumber\\
    \hat{W}_{j} ~\Big\{
    \ldots \otimes |t_{j-1}\rangle \otimes 
    |s_j\rangle \otimes 
    |t_{j+1}\rangle \otimes \ldots \Big\}
    &=&
    \Big\{
    \ldots \otimes|s_{j-1}\rangle \otimes 
    |s_j\rangle \otimes 
    |s_{j+1}\rangle
    \otimes \ldots \Big\}, \nonumber\\
    \hat{W}_{j} ~\Big\{
    \ldots \otimes |s_{j-1}\rangle \otimes 
    |t_j\rangle \otimes 
    |s_{j+1}\rangle \otimes \ldots \Big\}
    &=& (-)
    \Big\{
    \ldots \otimes|t_{j-1}\rangle \otimes 
    |t_j\rangle \otimes 
    |t_{j+1}\rangle
    \otimes \ldots \Big\}, \nonumber\\
    \hat{W}_{j} ~\Big\{
    \ldots \otimes |t_{j-1}\rangle \otimes 
    |t_j\rangle \otimes 
    |s_{j+1}\rangle \otimes \ldots \Big\}
    &=& (-)
    \Big\{
    \ldots \otimes|s_{j-1}\rangle \otimes 
    |t_j\rangle \otimes 
    |t_{j+1}\rangle
    \otimes \ldots \Big\}, \nonumber\\
    \hat{W}_{j} ~\Big\{
    \ldots \otimes |s_{j-1}\rangle \otimes 
    |t_j\rangle \otimes 
    |t_{j+1}\rangle \otimes \ldots \Big\}
    &=&(-)
    \Big\{
    \ldots \otimes|t_{j-1}\rangle \otimes 
    |t_j\rangle \otimes 
    |s_{j+1}\rangle
    \otimes \ldots \Big\}, \nonumber\\
    \hat{W}_{j} ~\Big\{
    \ldots \otimes |t_{j-1}\rangle \otimes 
    |t_j\rangle \otimes 
    |t_{j+1}\rangle \otimes \ldots \Big\}
    &=&(-)
    \Big\{
    \ldots \otimes|s_{j-1}\rangle \otimes 
    |t_j\rangle \otimes 
    |s_{j+1}\rangle
    \otimes \ldots \Big\}.
\end{eqnarray}

The kets $\vert s_j \rangle$ and $\vert t_j \rangle$ represent the two possibilities for $\vert \chi_j \rangle$.
The ket $\vert t_j \rangle$ should be interpreted as $\vert t_x \rangle$ on X bonds and as $\vert t_y \rangle$ on Y bonds. In these expressions, farther bond states are the same on the left and the right sides. For example, the bond state $\vert \chi_{j+2}\rangle$ is unaltered by the action of $\hat{W}_j$.
These relations can be gathered into a single expression,
\begin{eqnarray}
     \hat{W}_j  \Big\{
        \vert \chi_1 \rangle \otimes \ldots \otimes \vert \chi_N \rangle\Big\} 
        = g(\chi_j)
        \Big\{\ldots \otimes \vert {\chi}_{j-2}\rangle\otimes \vert \bar{\chi}_{j-1}\rangle \otimes \vert \chi_j \rangle \otimes \vert \bar{\chi}_{j+1}\rangle\otimes \vert {\chi}_{j+2}\rangle\otimes \ldots \Big\},
\end{eqnarray}
where $\bar{\chi}$ denotes a spin-flip and $g(\chi_j)=+1$ if $\chi_j$ is a singlet and $-1$ if $\chi_j$ is a triplet. Inserting this expression in Eq.~\ref{eq.WPchi}, we see that 
\begin{eqnarray}
\nonumber     \hat{W}_j \vert \chi_1,\ldots,\chi_N\rangle = 
    g(\chi_j) ~\Big\{
        \hat{P}^{S=1}_1\otimes \ldots\otimes  \hat{P}^{S=1}_N 
    \Big\}~
        \Big\{ \vert \chi_1 \rangle \otimes \ldots \vert \bar{\chi}_{j-1}\rangle \otimes \vert \chi_j\rangle \otimes \vert \bar{\chi}_{j+1}\rangle\ldots \otimes \vert \chi_N\rangle \Big\} \\
        = g(\chi_j) ~\vert \chi_1,\ldots,\bar{\chi}_{j-1},\chi_j,\bar{\chi}_{j+1},\ldots,\chi_N\rangle.
        \label{eq.WPchi2}
\end{eqnarray}  
We have arrived at Eq.~7 of the main text. We emphasize that the fractionalized states of Fig.~1 are not eigenstates of the $\hat{W}$ operators. At the same time, the $\hat{W}$ operators connect fractionalized states to one another, without taking them out of the fractionalized space.

\section{Spin-$\frac{1}{2}$ $XZX$ or Cluster model}
In the main text, the selection Hamiltonian of Eq.~9  acts on the Hilbert space of $N$ spin-$\frac{1}{2}$ degrees of freedom. The degrees of freedom are bond variables in the fractionalized construction. This Hamiltonian is, in fact, solvable by a Jordan-Wigner transformation\cite{yanagihara2020exact, sadaf2025quantum}. To write the ground-state wave function, we start with the Hamiltonian of Eq.~9 with all $w$'s set to $+1$ \cite{Tasaki2020,Wei2022,suzuki1971relationship,raussendorf2001one, keating2004random,son2011quantum, PhysRevB.96.165124, yanagihara2020exact, sadaf2025quantum},

\begin{eqnarray}
    \hat{H}_{(w_1=w_2=\ldots=w_N=+1)}= -\sum_{k=1}^N \sigma^x_{k-1}\sigma^z_{k}\sigma^x_{k+1}.
    \label{eq.HCluster}
\end{eqnarray}
Here, the $\sigma^{x,z}$ are Pauli matrices. This is the well-known cluster model. The exact ground state can be written down in the spin basis and expressed using a $2\times2$ MPS representation, see for example Ref. \cite{Tasaki2020}. The MPS representation is given by Fig.~3(b) where all $w$ indices are set to $+1$ (translationally invariant). The MPS matrices are given by $A^{\chi=\uparrow}_{w=+1}$ and $A^{\chi=\downarrow}_{w=+1}$ of Eq.~10.

We next seek to map Eq.~\ref{eq.HCluster} to Eq.~9 where $\{w_1,\ldots,w_N\}$ take arbitrary values, with each $w$ index being $\pm 1$. 
We identify the $w$ variables that are to be flipped from $+1$ to $-1$, say at $w_{k_1}$, $w_{k_2}$, etc. At each such location (each $k_i$), we perform a local $\pi$-rotation about $\hat{x}$ so that $\sigma_{k_i}^x$ is unchanged, but $\sigma_{k_i}^z \rightarrow (-)\sigma_{k_i}^z$. 
This transformation maps Eq.~\ref{eq.HCluster} to Eq.~9 of the main text with the desired $\{w\}$ values. 
It follows that the transformation also maps the ground state of Eq.~\ref{eq.HCluster} to that with the desired $\{w\}$ values. 

Since the unitary transformation flips the spin along the $\hat{z}$ direction and the MPS representation is in the $\sigma_z$ basis, we readily see that $A^{\chi=\uparrow}_{w=+1}=A^{\chi=\downarrow}_{w=-1}$ and $A^{\chi=\downarrow}_{w=+1}=A^{\chi=\uparrow}_{w=-1}$. We obtain the MPS form in Fig.~3(b) as the ground state of the selection Hamiltonian. The MPS representation is no longer translationally invariant (unless all $w$'s are set to -1) after the local unitary transformations are absorbed into the MPS matrices.

\section{The purely biquadratic model}
We consider the Hamiltonian of Eq.~1 with $K$, the strength of linear couplings, set to zero. 
\subsection{ $\theta=\frac{\pi}{2}$} 
At $\theta=\frac{\pi}{2}$, the Hamiltonian is given by
\begin{equation}
\hat{H}_{\theta=\frac{\pi}{2}}=\sum_j \left [
\left(\hat{S}_{2j}^x \hat{S}_{2j+1}^x\right)^2 + 
\left(\hat{S}_{2j+1}^y \hat{S}_{2j+2}^y \right)^2 \right ].
\end{equation}
As $(\hat{S}^x)^2$ and $(\hat{S}^y)^2$ are positive semi-definite operators, $H_{\theta=\frac{\pi}{2}}$ is a sum of positive semi-definite terms. Thus, the ground state energy has a lower bound of zero. If we are able to construct a state with energy zero, it must be a ground state. For periodic boundary conditions, the two product states depicted in Fig.~2(a) are readily seen to be annihilated by $\hat{H}_{\theta=\frac{\pi}{2}}$. In either state, every bond has one site in the $|S_x=0\rangle$ state and one site in the $|S_y=0\rangle$ state. The former ensures that the $(\hat{S}_{2j}^x \hat{S}_{2j+1}^x)^2$ term contributes zero, while the latter ensures that the $(\hat{S}_{2j}^y \hat{S}_{2j+1}^y)^2$ vanishes. With open boundary conditions, multiple arrangements of $|S_x=0\rangle$, $|S_y=0\rangle$ and $|S_z=0\rangle$ states produce zero-energy eigenstates. Exact diagonalization and combinatorial arguments show that the ground state degeneracy is $2N+1$.

With periodic boundaries, the two direct-product states of Fig.~2(a) continue to be ground states when a subdominant linear term is introduced. For $\frac{\pi}{4} \leq \theta \leq\frac{\pi}{2}$, the Hamiltonian may be rewritten as
\begin{equation}
    \hat{H}_{\frac{\pi}{4}\leq\theta\leq\frac{\pi}{2}}=\sqrt{2}~\cos \theta ~\hat{H}_{\theta=\frac{\pi}{4}} + (\sin \theta - \cos \theta) ~\hat{H}_{\theta=\frac{\pi}{2}}.
\end{equation}
Here $\hat{H}_{\theta=\frac{\pi}{4}}$ is defined in Eq.~3 and $(\sin \theta - \cos \theta)\geq0$ for this parameter regime. Thus, $\hat{H}_{\frac{\pi}{4}\leq\theta\leq\frac{\pi}{2}}$ is a sum of projectors and positive semi-definite operators. Any zero-energy eigenvectors are immediately seen to be ground states. With periodic boundary conditions, the two product states of Fig.~2(a) fit this requirement. 

\subsection{$\theta=\frac{3\pi}{2}$}

At $\theta=\frac{3\pi}{2}$, the Hamiltonian is given by
\begin{equation}
\hat{H}_{\theta=\frac{3\pi}{2}}=-\sum_j \left [
\left(\hat{S}_{2j}^x \hat{S}_{2j+1}^x\right)^2 + 
\left(\hat{S}_{2j+1}^y \hat{S}_{2j+2}^y \right)^2 \right ].
\end{equation}
The expectation value of each term in the Hamiltonian has a lower bound of $-1$. Any ansatz that saturates this bound must be a ground state. This is achieved by Fig.~2(b). At each site, the $|S_z=0\rangle=\frac{1}{\sqrt{2}}(|S_x=+1\rangle-|S_x=-1\rangle)=\frac{1}{i\sqrt{2}}(|S_y=+1\rangle-|S_y=-1\rangle)$ state is an eigenvector of both $(\hat{S}^x)^2$ and $(\hat{S}^y)^2$ with eigenvalue $+1$. Placing every site in the $|S_z=0\rangle$ state minimizes every term in the Hamiltonian and is therefore a ground state. 

With open boundary conditions, we construct the ground state in a similar way by placing $|S_z=0\rangle$ on all sites in the bulk of the chain. Spins at the edges have a two-fold choice. With $N$ even and edge bonds of the X type, each edge spin could be in either $|S_y=0\rangle=\frac{1}{\sqrt{2}}(|S_x=+1\rangle+|S_x=-1\rangle)$ or $|S_z=0\rangle$. The ground state has a four-fold degeneracy from this construction. 
The ground state energy comes out to be $-(N-1)$ for an $ N$- site chain with open boundaries, saturating the lower bound on energy. 

As seen from the phase diagram in Fig.~4, the $\theta=\frac{3\pi}{2}$ point is adiabatically connected to the pure Kitaev limit of $\theta=0$. In the case of the Kitaev chain with open boundaries, it is known that the ground state is four-fold degenerate\cite{sen2010spin,gordon2022insights,raja2025spin}. Our arguments here are consistent with this result. The four-fold degeneracy can also be interpreted in terms of the bond conserved quantities. In the bulk, all bonds have $w=+1$. The edge bonds could have $w=\pm 1$ independently. Placing an edge site in $\vert S_z=0\rangle$ or $\vert S_y=0\rangle$ leads to $w=+1$ or $w=-1$, respectively, on the edge bond. The four-fold degeneracy here is qualitatively different from that of the AKLT state. Here, the ground states can be constructed explicitly as direct-product states. In the AKLT model, a four-fold degeneracy is on account of a fractionalized dangling spin\cite{AKLT1987,affleck1988valence,Tasaki2020}. 

\section{MPS matrices for the ground states at $\theta=\frac{\pi}{4}$}
\begin{figure}
    \centering
\includegraphics[width=0.4\linewidth]{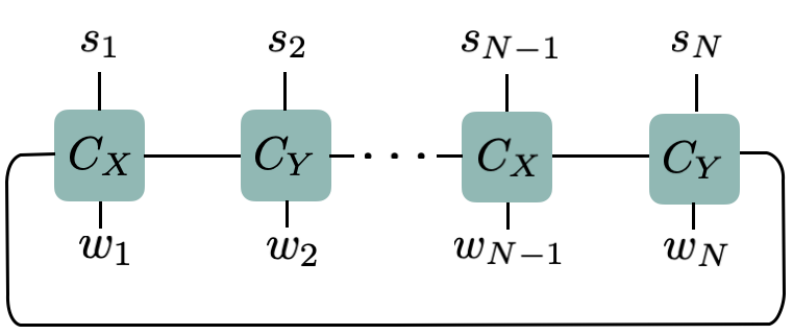}
    \caption{Matrices in Fig.~3(c) of the main text can be contracted to construct an MPS of bond dimension four as shown here. The $\{w\}$ indices represent bond conserved quantities, while $\{s\}$ indices are spin quantum numbers at each site.   }
    \label{fig.MPSCs}
\end{figure}
The ground states at the exactly solvable $\theta=\frac{\pi}{4}$ point are shown in Fig.~3(c). They are MPSs composed of $2\times 2$ matrices. 
They can be reexpressed as MPSs of bond dimension four by contracting matrices suitably, as shown in Fig.~\ref{fig.MPSCs}. Below, we give explicit expressions for the $4\times 4$ matrices involved. We first contract pairs of matrices,
\begin{eqnarray}
    B_X^{+1,\uparrow}&=&M^{+1}B_X^\uparrow=\frac{1}{\sqrt{2}}\begin{pmatrix}
        0 & 1\\
        0 & 0
    \end{pmatrix}=B_Y^{+1,\uparrow}, ~~~~~~~~~
    B_X^{-1,\uparrow}=M^{-1}B_X^\uparrow=\frac{1}{\sqrt{2}}\begin{pmatrix}
        0 & 0\\
        -1 & 0
    \end{pmatrix}=B_Y^{-1,\uparrow},\nonumber \\ 
        B_X^{0,\uparrow}&=&M^{+1}B_X^\uparrow=\frac{1}{2}\begin{pmatrix}
        -1 & 0\\
        0 & 1
    \end{pmatrix}=B_Y^{0,\uparrow},~~~~~~~~~~~
    B_X^{+1,\downarrow}=M^{+1}B_X^\downarrow=\frac{1}{\sqrt{2}}\begin{pmatrix}
        1 & 0\\
        0 & 0
    \end{pmatrix},\nonumber \\
    B_X^{-1,\downarrow}&=&M^{-1}B_X^\downarrow=\frac{1}{\sqrt{2}}\begin{pmatrix}
        0 & 0\\
        0 & -1
    \end{pmatrix},~~~~~~~~~~~~~~~~~
        B_X^{0,\downarrow}=M^{+1}B_X^\downarrow=\frac{1}{2}\begin{pmatrix}
        0 & -1\\
        1 & 0
    \end{pmatrix},\nonumber \\    
    B_Y^{+1,\downarrow}&=&M^{+1}B_Y^\downarrow=\frac{i}{\sqrt{2}}\begin{pmatrix}
        1 & 0\\
        0 & 0
    \end{pmatrix},~~~~~~~~~~~~~~~~~~~~
    B_Y^{-1,\downarrow}=M^{-1}B_Y^\downarrow=\frac{i}{\sqrt{2}}\begin{pmatrix}
        0 & 0\\
        0 & 1
    \end{pmatrix},\nonumber \\ 
        B_Y^{0,\downarrow}&=&M^{+1}B_Y^\downarrow=\frac{i}{2}\begin{pmatrix}
        0 & 1\\
        1 & 0
    \end{pmatrix}.      
\end{eqnarray}
Using these as building blocks, we obtain $4\times 4$ matrices. We first list matrices with bond conserved-quantity $w$ set to $+1$:
\begin{eqnarray}
    C_{X,w=+1}^{+1}&=&B_X^{+1,\uparrow} \otimes A^\uparrow_{w=+1}+B_X^{+1,\downarrow} \otimes A^\downarrow_{w=+1}  
    =\frac{1}{2}\begin{pmatrix}
        1 & 1 & 1 & 1\\
        -1 & 1 & 1 & -1\\
        0 & 0 & 0 & 0\\
        0 & 0 & 0 & 0
    \end{pmatrix}, \nonumber \\
     C_{Y,w=+1}^{+1}&=&B_Y^{+1,\uparrow} \otimes A^\uparrow_{w=+1}+B_Y^{+1,\downarrow} \otimes A^\downarrow_{w=+1} 
    =\frac{1}{2}\begin{pmatrix}
        i& i & 1 & 1\\
        -i & i & 1 & -1\\
        0 & 0 & 0 & 0\\
        0 & 0 & 0 & 0
    \end{pmatrix}, \nonumber \\
    C_{X,w=+1}^{-1}&=&B_X^{-1,\uparrow} \otimes A^\uparrow_{w=+1}+B_X^{-1,\downarrow} \otimes A^\downarrow_{w=+1} 
    =\frac{-1}{2}\begin{pmatrix}
        0 & 0 & 0 & 0\\
        0 & 0 & 0 & 0\\
        1 & 1 & 1 & 1\\
        1 & -1 & -1 & 1
    \end{pmatrix}, \nonumber \\     
    C_{Y,w=+1}^{-1}&=&B_Y^{-1,\uparrow} \otimes A^\uparrow_{w=+1}+B_Y^{-1,\downarrow} \otimes A^\downarrow_{w=+1}  
    =\frac{1}{2}\begin{pmatrix}
        0 & 0 & 0 & 0\\
        0 & 0 & 0 & 0\\
        -1 & -1 & i & i\\
        -1 & 1 & -i & i
    \end{pmatrix},  \nonumber \\  
    C_{X,w=+1}^0&=&B_X^{0,\uparrow} \otimes A^\uparrow_{w=+1}+B_X^{0,\downarrow} \otimes A^\downarrow_{w=+1} 
    =\frac{1}{2\sqrt{2}}\begin{pmatrix}
        -1 & -1 & -1 & -1\\
        -1 & 1 & 1 & -1\\
        1 & 1 & 1 & 1\\
        -1 & 1 & 1 & -1
    \end{pmatrix}, \nonumber \\     
    C_{Y,w=+1}^0&=&B_Y^{0,\uparrow} \otimes A^\uparrow_{w=+1}+B_Y^{0,\downarrow} \otimes A^\downarrow_{w=+1} 
    =\frac{1}{2\sqrt{2}}\begin{pmatrix}
        -1 & -1 & i& i\\
        -1 & 1 & -i & i\\
        i & i & 1 & 1\\
        -i & i & 1 & -1 
    \end{pmatrix}.   
\end{eqnarray}
We next list matrices with $w$ set to $-1$:
\begin{eqnarray}
    C_{X,w=-1}^{+1}&=&B_X^{+1,\uparrow} \otimes A^\uparrow_{w=-1}+B_X^{+1,\downarrow} \otimes A^\downarrow_{w=-1} 
    =\frac{1}{2}\begin{pmatrix}
        1 & 1 & 1 & 1\\
        1 & -1 & -1 & 1\\
        0 & 0 & 0 & 0\\
        0 & 0 & 0 & 0
    \end{pmatrix}, \nonumber \\
     C_{Y,w=-1}^{+1}&=&B_Y^{+1,\uparrow} \otimes A^\uparrow_{w=-1}+B_Y^{+1,\downarrow} \otimes A^\downarrow_{w=-1} 
    =\frac{1}{2}\begin{pmatrix}
        i& i & 1 & 1\\
        i & -i & -1 & 1\\
        0 & 0 & 0 & 0\\
        0 & 0 & 0 & 0
    \end{pmatrix}, \nonumber \\
    C_{X,w=-1}^{-1}&=&B_X^{-1,\uparrow} \otimes A^\uparrow_{w=-1}+B_X^{-1,\downarrow} \otimes A^\downarrow_{w=-1} 
    =\frac{-1}{2}\begin{pmatrix}
        0 & 0 & 0 & 0\\
        0 & 0 & 0 & 0\\
        1 & 1 & 1 & 1\\
        -1 & 1 & 1 & -1
    \end{pmatrix}, \nonumber \\     
    C_{Y,w=-1}^{-1}&=&B_Y^{-1,\uparrow} \otimes A^\uparrow_{w=-1}+B_Y^{-1,\downarrow} \otimes A^\downarrow_{w=-1}
    =\frac{1}{2}\begin{pmatrix}
        0 & 0 & 0 & 0\\
        0 & 0 & 0 & 0\\
        -1 & -1 & i & i\\
        1 & -1 & i & -i
    \end{pmatrix},  \nonumber \\  
    C_{X,w=-1}^0&=&B_X^{0,\uparrow} \otimes A^\uparrow_{w=-1}+B_X^{0,\downarrow} \otimes A^\downarrow_{w=-1} 
    =\frac{1}{2\sqrt{2}}\begin{pmatrix}
        -1 & -1 & -1 & -1\\
        1 & -1 & -1 & 1\\
        1 & 1 & 1 & 1\\
        1 & -1 & -1 & 1
    \end{pmatrix}, \nonumber \\     
    C_{Y,w=-1}^0&=&B_Y^{0,\uparrow} \otimes A^\uparrow_{w=-1}+B_Y^{0,\downarrow} \otimes A^\downarrow_{w=-1} 
    =\frac{1}{2\sqrt{2}}\begin{pmatrix}
        -1 & -1 & i& i\\
        1 & -1 & i & -i\\
        i & i & 1 & 1\\
        i & -i & -1 & 1 
    \end{pmatrix}.   
    \label{eq.w-1matrices}
\end{eqnarray}
These matrices yield unnormalized MPS states. They should be normalized before evaluating any expectation values. The norm, $\mathcal{N}$, can be evaluated as
\begin{eqnarray}
    \tilde{C}_{X,w}&=&(C_{X,w}^{+1})^*\otimes C_{X,w}^{+1} + (C_{X,w}^{-1})^*\otimes C_{X,w}^{-1} + (C_{X,w}^0)^*\otimes C_{X,w}^0 , ~~~~~~\nonumber \\
    \tilde{C}_{Y,w}&=&(C_{Y,w}^{+1})^*\otimes C_{Y,w}^{+1} + (C_{Y,w}^{-1})^*\otimes C_{Y,w}^{-1} + (C_{Y,w}^0)^*\otimes C_{Y,w}^0, \nonumber \\
    \mathcal{N}&=& \Big\{\mathrm{Tr}(\tilde{C}_{X,w}\tilde{C}_{Y,w}....\tilde{C}_{X,w}\tilde{C}_{Y,w})\Big\}^{\frac{1}{2}}.
\end{eqnarray}

\section{Fractionalized variational ansatz for $\theta<\frac{\pi}{4}$: Uniform $w=+1$ state }
The phase diagram of Fig.~4 exhibits a unique ground state for $\theta > \frac{3\pi}{4}$ and $\theta <\frac{\pi}{4}$. In terms of the bond conserved quantities, this state has all $w$'s set to $+1$. Here, we seek to describe certain qualitative features of this wave function. This state can be approximated as a linear superposition of fractionalized states of the form shown in Fig.~1. The bond conserved quantities place strong constraints on the form of the linear superposition. To see this, we consider three global spin rotations by $\pi$: about the X, Y and Z axes. We express these rotations as operators,

\begin{eqnarray}
    \hat{U}^x=\prod_i^N\exp(i\pi \hat{S}_i^x)=\prod_j \hat{W}_{2j}, ~~~
    \hat{U}^y=\prod_i^N\exp(i\pi \hat{S}_i^y)=\prod_j \hat{W}_{2j+1}, ~~~
    \hat{U}^z=\prod_i^N\exp(i\pi \hat{S}_i^z)=\prod_j \hat{W}_j.
\end{eqnarray}
In each expression, the index $j$ runs over all integers. As seen from these expressions, 
$\hat{U}^x \hat{U}^y = \hat{U}^z$.
Any state in the uniform $w=+1$ sector is readily seen to be invariant under each of the three rotations. In particular, this applies to the ground state in the vicinity of $\theta=0$.
As this state is adiabatically connected to the exactly solvable point at $\theta=\frac{\pi}{4}$, the ground state can be well approximated as a linear superposition of fractionalized states.

We consider the action of the rotation operators on each fractionalized state of the form in Fig.~1. The singlet bond state is invariant under any rotation. The triplet-x state is invariant under $\hat{U}^x$, but switches sign under $\hat{U}^y$ and $\hat{U}^z$. Similarly, the triplet-y state is invariant under $\hat{U}^y$, but switches sign under $\hat{U}^x$ and $\hat{U}^z$. For the $\theta=0$ ground state to be invariant under $\hat{U}^x$, the number of triplet-y bond states must be even. To be invariant under $\hat{U}^y$, the number of triplet-x bond states must be even. This reveals a hidden structure in the $\theta=0$ ground state. It is composed of fractionalized states that are subject to global constraints: on X (Y) bonds, we must have an even number of triplet-x (triplet-y) states.

\section{Direct Product States within the fractionalized set: Uniform $w=-1$ state}
We describe the relationship between the direct-product states of Fig.~2(a) and the fractionalized states of Fig.~1. We first characterize the direct-product states in terms of the bond conserved quantities. The rotation operator $e^{i\pi \hat{S}_j^x}$ leaves $\vert S_x=0\rangle$ invariant, but changes the sign of $\vert S_y=0\rangle$. Similarly, $e^{i\pi \hat{S}_j^y}$ changes the sign of $\vert S_x=0\rangle$ but leaves $\vert S_y=0\rangle$ unchanged. From these properties, we see that the direct-product states of Fig.~2(a) have $w=-1$ on every bond. 
\begin{figure}
    \centering
\includegraphics[width=0.7\linewidth]{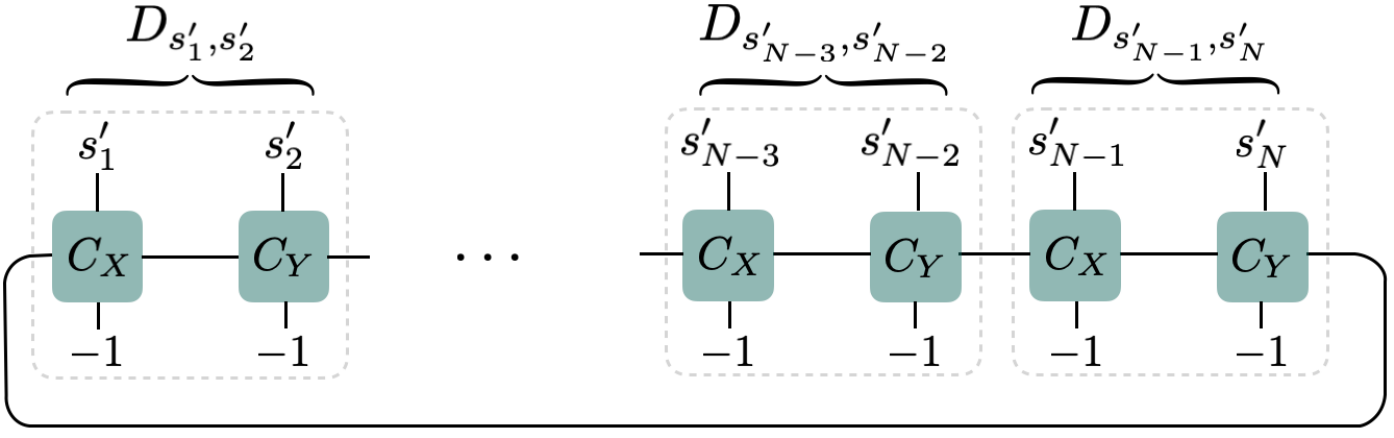}
    \caption{The fractionalized ground state with all $w$'s set to $-1$. The local spin indices are given in the $s'\in \{x,y,z\}$ basis. Pairs of matrices can be contracted into $4\times 4$ matrices of the form $D_{\alpha\beta}$, where $\alpha,\beta=x,y,z$.  }
    \label{fig.Ds}
\end{figure}
We next explore whether the direct product states can be written in terms of the fractionalized states of Fig.~1. 
We consider the MPS representation of Fig.~3(c) with all $w$'s set to $-1$. Below, we demonstrate that this MPS state is precisely the symmetric combination of the two direct-product states. 

In Fig.~3(c), the MPS is expressed in terms of local spin-z quantum numbers, $s=-1,0,1$, representing three states of a spin-1 moment at each site. For convenience, we switch to a basis given by $\{|S_x=0\rangle,|S_y=0\rangle,|S_z=0\rangle\}$, labeled for brevity as $s'=x,y,z$. With this basis change, the matrices of Eq.~\ref{eq.w-1matrices} transform as
\begin{eqnarray}
    C_{X,w=-1}^{s'=z}&=&C_{X,w=-1}^{0} , ~~~~~~~~~~~~~~~~~~~~~~~~~~~~~~~~~~~~~~~
    C_{Y,w=-1}^{s'=z}=C_{Y,w=-1}^{s=0}, \nonumber \\
    C_{X,w=-1}^{s'=x}&=&\frac{1}{\sqrt{2}}(C_{X,w=-1}^{s=+1} -C_{X,w=-1}^{s=-1}), ~~~~~~~~~~~~~~~~
    C_{Y,w=-1}^{s'=x}=\frac{1}{\sqrt{2}}(C_{Y,w=-1}^{s=+1} -C_{Y,w=-1}^{s=-1}),\nonumber \\
    C_{X,w=-1}^{s'=y}&=&\frac{1}{\sqrt{2}}(C_{X,w=-1}^{s=+1} +C_{X,w=-1}^{s=-1}), ~~~~~~~~~~~~~~~~
    C_{Y,w=-1}^{s'=y}=\frac{1}{\sqrt{2}}(C_{Y,w=-1}^{s=+1} +C_{Y,w=-1}^{s=-1}).
\end{eqnarray}    
With $N$ spins on a periodic chain, the MPS representation contains an alternating sequence of $C_{X,w=-1}^{\alpha}$ and $C_{Y,w=-1}^{\beta}$ matrices, where $\alpha,\beta=x,y,z$. 
We contract pairs to write
\begin{eqnarray}
    D_{\alpha\beta}=C_{X,w=-1}^{\alpha}C_{Y,w=-1}^{\beta},
\end{eqnarray}
as shown in Fig.~\ref{fig.Ds}. With three choices for $\alpha$ and three for $\beta$, we have nine distinct $D$ matrices. 

\begin{figure}
    \centering
\includegraphics[width=0.4\linewidth]{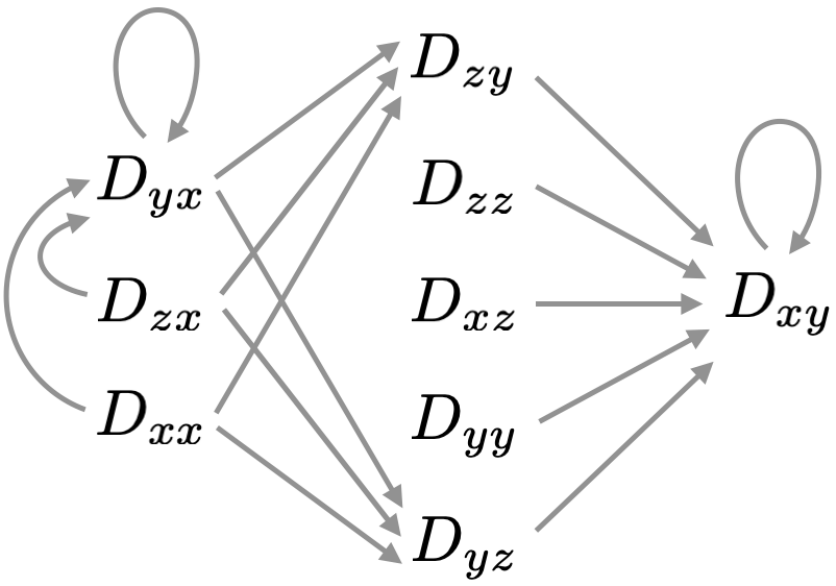}
    \caption{Multiplicative structure of $D$ matrices. Each arrow indicates a pair of matrices (in order) that can be multiplied to yield a non-zero result. To have a product of $\frac{N}{2}$ matrices with non-zero trace, we must follow a closed path on this network, with each segment traversed in the direction of the arrow. Only two such closed paths are possible: choosing all matrices to be $D_{xy}$ or choosing all matrices to be $D_{yx}$.}
    \label{fig.Darrows}
\end{figure}

The MPS representation involves a product of $\frac{N}{2}$ $D_{\alpha\beta}$ matrices. We next consider the product of a pair of $D_{\alpha\beta}$ matrices. With nine possibilities for $D_{\alpha\beta}$, we have 81 such products. However, only 15 of them are non-zero. For example, $D_{yx}D_{xy}$ is zero (all elements vanish), while $D_{yx}D_{zy}$ has non-zero elements. 
The non-zero products are depicted in Fig.~\ref{fig.Darrows} where each arrow shows a pair of matrices (in order) that can be multiplied to obtain a non-zero result. In the MPS, $\frac{N}{2}$ $D_{\alpha\beta}$ matrices should be multiplied to yield a non-zero trace. This is equivalent to traversing the `graph' in Fig.~\ref{fig.Darrows} to form closed loops, with each segment traversed in the direction of the corresponding arrow. There are only two possible ways to obtain a closed loop: $(D_{xy}D_{xy}D_{xy}....D_{xy})$ and $(D_{yx}D_{yx}D_{yx}...D_{yx})$. That is, all $\frac{N}{2}$ matrices should either be $D_{xy}$ or all should be $D_{yx}$.

We are left with two non-zero coefficients for the wave function of the fractionalized state with all $w$'s set to $-1$. They are given by $\mathrm{Tr}(D_{xy}D_{xy}D_{xy}....D_{xy})$ and $\mathrm{Tr}(D_{yx}D_{yx}D_{yx}...D_{yx})$. From the explicit forms of $D_{xy}$ and $D_{yx}$, these two are seen to be equal. These two coefficients correspond precisely to the two direct-product states in Fig.~2(a). This demonstrates that the fractionalized state in the uniform $w=-1$ sector is the symmetric combination of the product states in Fig.~2(a). In turn, this establishes that the antisymmetric combination of product states is orthogonal to the set of $2^N$ fractionalized states. This follows from the fact that each $\{w\}$ sector allows for one fractionalized state, given by Fig.~3(c). A second state with the same $\{w\}$ configuration must necessarily be orthogonal to all fractionalized states.

\section{Bond conserved quantities in the phase with the unique ground state}

We present an analytic argument as to why the ground state for $(\theta < \frac{\pi}{4},~\theta>\frac{3\pi}{4})$ is in the uniform $w=+1$ sector. We begin with an observation regarding the bond conserved quantities. As discussed above, the ground state at $\theta=\frac{3\pi}{2}$ is obtained by placing $|S_z=0\rangle$ on all sites. It lies in the sector where all bond conserved quantities are $+1$. If we were to replace a pair of $|S_z=0\rangle$'s on an X bond with a pair of $|S_y=0\rangle$'s, this does not alter the conserved quantities. Likewise, on a Y bond, we may replace the pair of $|S_z=0\rangle$'s with a pair of $|S_x=0\rangle$'s without changing the $w$'s.

To move away from the exactly solvable point at $\theta=\frac{3\pi}{2}$ in either direction, we may introduce bilinear terms as a small perturbation, i.e., with $\vert K \vert \ll 1$. We describe the action of the bilinear terms on the $\theta=\frac{3\pi}{2}$ ground state below:
\begin{enumerate}
    \item $\hat{S}_{2j}^x \hat{S}_{2j+1}^x$ acting once : $\hat{S}_{2j}^x \hat{S}_{2j+1}^x \left\{... |S_z=0\rangle_{2j}|S_z=0\rangle_{2j+1}...\right\} = \left\{... |S_y=0\rangle_{2j}|S_y=0\rangle_{2j+1}...\right\}$,
    \item $\hat{S}_{2j}^x \hat{S}_{2j+1}^x$ acting twice : $\hat{S}_{2j}^x \hat{S}_{2j+1}^x \left\{... |S_y=0\rangle_{2j}|S_y=0\rangle_{2j+1}...\right\} = \left\{... |S_z=0\rangle_{2j}|S_z=0\rangle_{2j+1}...\right\}$,
    \item $\hat{S}_{2j+1}^y \hat{S}_{2j+2}^y$ acting once : $\hat{S}_{2j+1}^y \hat{S}_{2j+2}^y \left\{... |S_z=0\rangle_{2j+1}|S_z=0\rangle_{2j+2}...\right\} = \left\{... |S_x=0\rangle_{2j+1}|S_x=0\rangle_{2j+2}...\right\}$,
    \item $\hat{S}_{2j+1}^y \hat{S}_{2j+2}^y$ acting twice : $\hat{S}_{2j+1}^y \hat{S}_{2j+2}^y \left\{... |S_x=0\rangle_{2j+1}|S_x=0\rangle_{2j+2}...\right\} = \left\{... |S_z=0\rangle_{2j+1}|S_z=0\rangle_{2j+2}...\right\}$,
    \item $\hat{S}_{2j}^x \hat{S}_{2j+1}^x$ acting once followed by $\hat{S}_{2j+1}^y \hat{S}_{2j+2}^y$ : $\hat{S}_{2j+1}^y \hat{S}_{2j+2}^y\left\{... |S_y=0\rangle_{2j}|S_y=0\rangle_{2j+1}...\right\} =0$,
    \item $\hat{S}_{2j+1}^y \hat{S}_{2j+2}^y$ acting once followed by $\hat{S}_{2j}^x \hat{S}_{2j+1}^x$ : $\hat{S}_{2j}^x \hat{S}_{2j+1}^x\left\{... |S_x=0\rangle_{2j+1}|S_x=0\rangle_{2j+2}...\right\} =0$.
\end{enumerate}
In each expression, the right-hand side is unnormalised and defined up to a global phase. Based on these relations, we conclude that: 
\begin{enumerate}
    \item The lowest-order energy correction is obtained at second order. 
    \item Intermediate states lie in the uniform $w=+1$ sector. They involve replacing pairs of $\vert S_z=0\rangle$'s on X bonds (Y bonds) with $\vert S_y=0\rangle$'s  ($\vert S_x=0\rangle$'s). This retains the uniform $+1$ value of bond conserved quantities.  
    \item Conserved quantities are unchanged at \textit{all} orders in perturbation theory. The lowest order corrections have $|S_y=0\rangle$'s placed on a single X bond or $|S_x=0\rangle$'s placed on a single Y bond, with $|S_z=0\rangle$ on the remaining $N-2$ sites. The next -order correction has either $|S_y=0\rangle$'s on any 2 X bonds, $|S_x=0\rangle$'s on any 2 Y bonds or 2 non-consecutive X and Y bonds with $|S_y=0\rangle$'s and $|S_x=0\rangle$'s respectively. This pattern in the order of perturbation theory continues until all uniform $w=+1$ states are accessed. 
    \item In the vicinity of $\theta=\frac{3\pi}{2}$, the state with $|S_z=0\rangle$ placed on all the sites has the highest weight in the ground state eigenvector, followed by the $N$ states with either $|S_y=0\rangle$'s placed on a single X bond or $|S_x=0\rangle$'s placed on a single Y bond.
    \item As the conserved quantities are unchanged at all orders in perturbation theory, we obtain an expansive region around $\theta=\frac{3\pi}{2}$ where the ground state is in the uniform $w=+1$ sector. 
\end{enumerate}

\section{Extending the model to trivalent graphs with $S=\frac{3}{2}$}

The extended spin-1 Kitaev model described in the main text naturally generalizes to higher spin values and different lattice geometries. As an illustrative example, we discuss the model for $S=\frac{3}{2}$. This requires a lattice (or cluster) with a coordination number that is a multiple of three, in order to accommodate a pair of spinons along each bond. We consider the simplest possibility -- a tetrahedron as shown in Fig.~6(a) of the main text.

In the main text, Eq.~3 describes a projector Hamiltonian for $S=1$. To construct the analogue for $S=\frac{3}{2}$, the Hamiltonian on the bond connecting sites $i$ and $j$ must be
\begin{eqnarray}
 \hat{P}_{ij}^{S^\alpha_{total}=\pm 3}
=\frac{1}{1152}(4S_i^\alpha S_j^\alpha+9)(4(S_i^\alpha)^2-1)(4(S_j^\alpha)^2-1) \nonumber\\
=\frac{1}{360}\Big[ 20\,{S_{i\alpha} }^3 \,{S_{j\alpha} }^3 -5\,{S_{i\alpha} }^3 \,S_{j\alpha} +45\,{S_{i\alpha} }^2 \,{S_{j\alpha} }^2 
-\frac{45\,({S_{i\alpha} }^2+{S_{j\alpha} }^2) }{4}-5\,S_{i\alpha} \,{S_{j\alpha} }^3 +\frac{5\,S_{i\alpha} \,S_{j\alpha} }{4}+\frac{45}{16}   \Big],
\end{eqnarray}
where $\alpha=x,y,z$ is the bond direction. Here, $\pm 3$ are the maximal values of the bond angular momentum along the bond direction.

To construct ground states, we fractionalize the $S=\frac{3}{2}$ moment at each vertex into three spinons. On each bond, we place the two spinons at either end into one of two bond states. On $X$ and $Y$ bonds, bond states are encoded by the matrices in Eq.~5. On Z bonds, we define
\begin{eqnarray}
    B_Z^\uparrow = \left(
    \begin{array}{cc}
     0 & \frac{1}{\sqrt{2}} \\
     -\frac{1}{\sqrt{2}} & 0
    \end{array}
    \right),~~~~
    B_Z^\downarrow = \left(
    \begin{array}{cc}
     0 & \frac{1}{\sqrt{2}} \\
     \frac{1}{\sqrt{2}} & 0
    \end{array}
    \right).
\end{eqnarray}
As with $B_X^\uparrow$ and $B_Y^\uparrow$, $B_Z^\uparrow$ denotes a singlet bond state. We introduce a triplet-z state, $\vert t_z\rangle = \frac{1}{\sqrt{2}} \left\{\vert \! \uparrow \downarrow \rangle + \vert \!\downarrow \uparrow \rangle \right\}$, represented by the matrix $B_Z^\downarrow$ above. On each site, we define a $M^s_{a,b,c}$ tensor that projects the three local spinons into the $S=\frac{3}{2}$ space.  Here, $s$ denotes the physical leg and takes $\pm \frac{3}{2}, \pm \frac{1}{2}$ values in the local $S^z$ basis of spin-$\frac{3}{2}$. We have $a,b,c=\pm\frac{1}{2}$, which are virtual indices that get contracted on each of the 3 bonds connected to a site. The non-zero elements of the tensor $M$ are obtained from the Clebsch-Gordan coefficients of adding three spin-$\frac{1}{2}$'s in the total spin-$\frac{3}{2}$ sector, and they are given below.
\begin{eqnarray}
    M^{+\frac{3}{2}}_{1/2,1/2,1/2} = 1, \nonumber\\
    M^{+\frac{1}{2}}_{-1/2,1/2,1/2} = M^{+\frac{1}{2}}_{1/2,-1/2,1/2}= M^{+\frac{1}{2}}_{1/2,1/2,-1/2}=1/\sqrt{3}, \nonumber\\
    M^{-\frac{1}{2}}_{-1/2,-1/2,1/2} = M^{-\frac{1}{2}}_{1/2,-1/2,-1/2}= M^{-\frac{1}{2}}_{-1/2,1/2,-1/2}=1/\sqrt{3}, \nonumber\\
    M^{-\frac{3}{2}}_{-1/2,-1/2,-1/2} = 1.
\end{eqnarray}

The tetrahedron, with six bonds, allows for $2^6=64$ fractionalized states. These states are constructed as PEPSs as shown in Fig.~6(b). They are ground states by construction, as they are annihilated by the projector Hamiltonian. We have explicitly constructed the 64 states and checked that they are linearly independent as well. By exact diagonalization, we find that the tetrahedron has 76 ground states. Apart from the fractionalized states, we have 12 additional ground states. Note that the tetrahedron geometry does not allow for bond conserved quantities. The operators defined in Eq.~2 no longer commute with the Hamiltonian, as each end of a bond is attached to two other `perpendicular' bonds. For example, we cannot define a $\hat{W}$ operator on an X bond that commutes with the adjacent Y and Z bonds simultaneously. We may define conserved quantities on plaquettes. However, the number of plaquette-conserved-quantities is much less than the number of fractionalized states. As a result, they do not suffice to orthogonalize the fractionalized states.

The $S=\frac{3}{2}$ problem can also be defined on a ladder geometry, as shown in Fig.~6(c). We have explicitly constructed the ground states for a four-rung ladder (with eight sites) with periodic boundaries. With twelve bonds, we obtain $2^{12}=4096$ fractionalized states -- ground states by construction. We have checked, by explicit construction, that these states are linearly independent. 

\end{document}